\documentclass[journal]{vgtc}                        
\usepackage{mathptmx}
\usepackage{graphicx}
\usepackage{times}
\usepackage{color}
\usepackage{microtype}                 
\PassOptionsToPackage{warn}{textcomp} 
\usepackage{textcomp}                 
\usepackage{mathptmx}                 
\usepackage{times}                   
        
\usepackage{cite}
\usepackage{fancyhdr}
\definecolor{blue}{rgb}{0,0,0}
\definecolor{red}{rgb}{0,0,0}
\definecolor{green}{rgb}{0,0,0}

\onlineid{0}
\vgtccategory{Research}
\vgtcpapertype{please specify}

\title{Force Rendering and Its Evaluation of \\a Friction-based Walking
Sensation Display for a Seated User}

\author{
Ginga Kato\thanks{e-mail:ginga765@gmail.com}\\ %
        \parbox{1.4in}{\scriptsize \centering 
        Graduate School of Information Science and Technology,\\ Osaka University}
\and 
Yoshihiro Kuroda\thanks{e-mail:ykuroda@bpe.es.osaka-u.ac.jp }\\ %
     \parbox{1.4in}{\scriptsize \centering 
     Graduate School of Engineering Science,\\ Osaka University} %
\and 
Kiyoshi Kiyokawa\thanks{e-mail:kiyo@is.naist.jp}\\ %
     \parbox{1.4in}{\scriptsize \centering 
     Graduate School of Information Science,\\ Nara Institute of Science and Technology}
\and 
Haruo Takemura\thanks{e-mail:takemura@cmc.osaka-u.ac.jp}\\ %
     \parbox{1.4in}{\scriptsize \centering 
     Cyberme
      Center,\\ Osaka University}
     }

\abstract{
{\color{green}Most existing locomotion devices that
represent the sensation of walking
target a user who is actually performing a walking motion.}
{\color{blue} Here, we attempted} to represent the walking sensation, {\color{red}especially a kinesthetic sensation and advancing feeling}
{\color{green}(the sense of moving forward)}
while the user remains seated. 
{\color{blue}To represent the walking sensation using a relatively simple device, we} focused on the force rendering and its evaluation of the longitudinal friction force applied on the sole 
during walking.
Based on the measurement of the friction force applied on the sole during actual walking, we developed a novel friction force display 
that can present the friction force without the influence of body weight.
{\color{blue}Using performance evaluation testing, 
we found} that 
the proposed method {\color{blue}can stably and rapidly} display friction force. 
  {\color{blue}Also}, we developed a virtual reality (VR) walk-through system that {\color{blue}is able to} present the friction force {\color{blue}through} 
the proposed device according to the avatar's 
walking motion in {\color{blue}a} virtual world. 
{\color{green}By evaluating the realism,}
we {\color{blue}found} that 
the proposed device can represent a more realistic advancing feeling {\color{blue}than} vibration feedback.
} 

\keywords{Virtual Reality, Locomotion Interface, Seated Position, Walking Sensation, Friction Force}

\CCScatlist{ 
  \CCScat{Hardware}{Haptic devices}{}{};
  \CCScat{Human-centered computing}{Virtual reality}{}{}
}

\begin{document}
\maketitle

\thispagestyle{fancy}

\section{Introduction}
Recently, {\color{blue}many} haptic devices representing realistic {\color{red}dexterous hand manipulation and body movement} 
have been developed.
Among these devices, some {\color{blue}attempt} to represent the sensation {\color{blue}of walking} to improve immersion in {\color{blue}the} virtual world.
Most of these locomotion devices represent the sensation {\color{blue}of} walking {\color{red}on a solid ground} in {\color{blue}an} infinite space 
by canceling the user's movement while the user is performing the {\color{blue}real} walking motion in a standing posture 
\cite{redirection1, redirection2, redirectionBlink, GaitMaster, TorusTreadmill, CyberCarpet, Treadport, VirtuixOmni, Fontana, Steinicke}. However, these devices tend to be {\color{blue}large} and {\color{blue}require} some support instruments. {\color{blue}Also, although the real} walking motion improves the {\color{red}realism} of walking, 
the motion also causes fatigue. 
{\color{blue}Therefore, {\color{red}to} enable users to enjoy virtual walking for a longer time, it is necessary to reduce {\color{blue}this} fatigue.}

{\color{blue}Here we attempt} to represent the walking sensation while the user remains seated.
In the case of a seated walking display, the user would feel less fatigue{\color{blue},} 
and the device does not {\color{blue}require additional}
 supports because the {\color{blue}real} walking motion is not involved. 
{\color{blue}Some} previous studies {\color{blue}have attempted to represent} the walking sensation to {\color{blue}a seated} user \cite{Pedal, HapticFootstepDisplay, Vibration, VirtualIsu}. 
However, these studies focused only on the representation of leg motion or the pressure and impact applied to the 
sole of the foot{\color{red}, rather than the sole of the shoe}.
{\color{blue}Few studies have attempted} to represent the longitudinal friction force on {\color{blue}the} sole generated as {\color{blue}a} 
reaction force {\color{blue}following kicking} the ground {\color{blue}during walking}. 

We thought that the friction force applied by kicking the ground is especially important {\color{blue}when representing} {\color{red}kinesthetic sensation} and the 
advancing feeling 
{\color{green}(the sense of moving forward)}
during walking. Therefore, {\color{blue}here we attempted} to represent the walking sensation by displaying this 
longitudinal friction force on the sole. 

\begin{figure}[tb]
  \centering
  \includegraphics[width=1.475in]{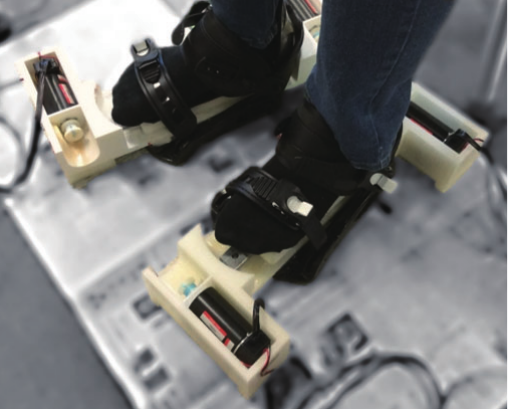}
  \includegraphics[width=1.8in]{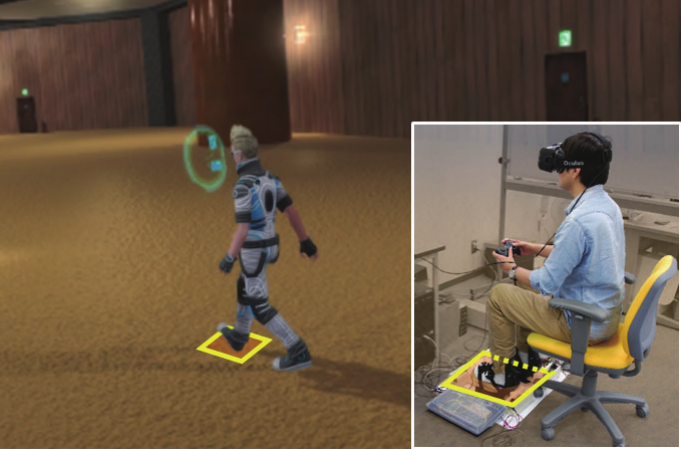}\\
  \hspace{-1mm}(a)\hspace{41mm}(b) \vspace{-3mm}
  \caption{(a) The friction-based walking sensation display considering the impulse by friction forces (b) example of use \label{TopHapStep}} \vspace{-5mm}
\end{figure}
 
\section{Related works}
\subsection{Locomotion Devices with {\color{blue}Real} Walking Motion}
{\color{blue}Many} locomotion devices {\color{blue}have been developed that aim at representing the sensation of walking in a virtual world}.
Many of these devices cancel the user's movement while the user is performing the {\color{blue}real} walking motion in a standing posture.

For example, 
\textcolor{red}{a redirection approach~\cite{redirection1, redirection2, redirectionBlink}} 
\textcolor{blue}{is an attempt} to extend \textcolor{blue}{a} limited walking space to an infinite space by secretly modifying the 
images displayed in the head\textcolor{blue}{-}mounted display (HMD) and guiding the user. Nevertheless, guidance with spatially distorted images \textcolor{blue}{cannot be used within} a small walking space 
\textcolor{green}{and thus requires} a large walking space.
\textcolor{blue}{Guidance} 
without discomfort
\textcolor{blue}{has been achieved} by modifying the images during \textcolor{blue}{blinking~\cite{redirectionBlink}.} 
However, \textcolor{blue}{this requires} a non-narrow walking space. 

On the other hand, some studies \textcolor{blue}{have attempted} to cancel the user's motion by moving the floor. 
If the floor is moved in the opposite direction of the user's movement, the user \textcolor{blue}{can} 
walk infinitely in a limited space. For example, \textcolor{blue}{using a parallel-link mechanism, Iwata et al. \cite{GaitMaster} move} the plate on which the user's feet are put.
{\color{blue}Also}, 
\textcolor{red}{several virtual walking systems \cite{TorusTreadmill, CyberCarpet, Treadport}} use a treadmill mechanism to maintain the user's position. 
{\color{blue}During real} walking, {\color{blue}we kick} the ground {\color{blue}with each step and, } 
as a reaction {\color{blue}to} the kicking force, {\color{blue}a} friction force is applied {\color{blue}to} the sole of the foot.
Such a friction force works as a driving force to advance the body forward.
However, {\color{blue}when using} devices that {\color{blue}move} the floor, 
the movement of the floor cancels 
the kicking force, and the user feels weaker friction on the floor. 
Thus, the advancing feeling produced by these approaches is limited.
Meanwhile, the device in \cite{VirtuixOmni} {\color{blue}attempted} to reduce the driving force itself, {\color{blue}which} is generated by kicking the 
ground, by reducing the friction between the sole of the shoes and the bowl-shaped walking space{\color{blue}; this} device needs no actuator {\color{blue}and, therefore,} the device {\color{blue}can be relatively compact}. 
However, these approaches have a common problem {\color{blue}in} that they cause the 
feeling {\color{blue}of} walking with {\color{blue}little} friction or on a slippery floor. 

{\color{red}Another idea is to make the user {\color{blue}take} steps on the spot \textcolor{red}{~\cite{WIP}. These authors attempted} to represent the walking sensation by making the user {\color{blue}perform a} stepping-like motion. In this method, only the sensor to detect the user's step is needed, and there {\color{blue}is} no need to prepare a large apparatus or space. 
{\color{blue}In addition to} the stepping-like motion, {\color{blue}an} arm swing motion can {\color{blue}also} generate {\color{blue}a} realistic walking sensation \cite{AS}. 
Kitson et al. compared a motion-cuing locomotion interface with a general joystick~\cite{Kitson}.
Zanotto et al. {\color{blue}have} used a vibrotactile feedback on foot using embedded sensors and actuators in the footwear~\cite{Audio}. 
Another paper~\cite{PSY} compared the walking interaction types and evaluated their naturalness. 
However, {\color{blue}although} those types of study {\color{blue}use 
cost-effective devices,} the advancing feeling is limited.}

When the user is performing the {\color{blue}real} walking motion {\color{red} or stepping-like motion}{\color{blue},} and the visibility is obscured by the HMD, 
there is a danger {\color{red} that users will lose their balance}. Thus, {\color{red} to provide a safe virtual walking experience, }
some support instruments are needed. 
{\color{red}Also, the real walking motion causes fatigue, making long-time experience difficult, although it cannot be simply regarded as a problem in terms of realism.}

\subsection{Walking Displays while the User Remains Seated}
Some previous studies {\color{blue}have attempted to produce a} walking sensation {\color{blue}to a seated user.} 
\textcolor{red}{A pedal-structured device has been used} to represent the walking sensation to {\color{blue}a seated} user by 
representing the motion of the leg during walking~\cite{Pedal}. 
\textcolor{red}{Using a mechanism of lifting plates, Ravikrishnan et al.~\cite{HapticFootstepDisplay} attempted} to represent the walking sensation 
by representing the pressure applied {\color{blue}to} the sole during walking. 
In the case of these ``seated" walking displays, the user's body is supported by the chair {\color{blue}and, thus, no additional supports are needed. Also}, these devices do not cause fatigue.
However, these devices need to be driven {\color{blue}by a strong actuator due to the weight of the user's legs}.

On the other hand, {\color{red}there are devices that represent the impact {\color{blue}applied to} the sole when the foot {\color{blue}is} landed 
by {\color{blue}giving a} vibration stimulus~\cite{Vibration0, Vibration}. Those} studies {\color{blue}were able to use a simple device because of a vibration actuator.} 
{\color{red}Terziman et al. added an artificial visual effect named "King-Kong effects" to enhance the sensation of walking~\cite{KingKong}.}
{\color{blue}Another study} utilized the footstep motion while sitting as an input to 
advance the avatar in the virtual world~\cite{VirtualIsu}.  
In that study, 
the footstep motion is utilized not only as an input but also as a ``manual actuator" 
to present the pressure on its own sole. The device needs no ``machine actuator", 
thus the compact design is realized.

{\color{blue}Although the devices mentioned above} represent leg motion, pressure, and impact during walking,
there are few studies that represent the friction force applied when kicking the ground.
{\color{blue}During real walking}, when the foot {\color{blue}is landed,} or the {\color{blue}walker kicks} the ground, {\color{blue}a} longitudinal friction force 
is applied {\color{blue}to} the sole. This friction force works as the brake and the driving force during walking.
Thus, the sensation represented by the previous devices without the friction force is similar to the sensation of footsteps on the spot, {\color{red}which does not represent the haptic feeling of advancing.} 
We assume this friction force is important to represent {\color{red}walking sensations involving the feeling of advancing.} 
{\color{blue}The total impulse by forward and backward friction forces is especially important for determining} the total advancement. 
Kato et al. demonstrated a prototype of presenting {\color{blue}this} friction force~\cite{GKato}. 
However, the displayed force considers {\color{blue}only} the timing of hitting and kicking the ground{\color{blue},} without considering the impulse by friction forces, and no evaluation was reported. 

{\color{blue}Here we} aimed to represent the sensation of walking by presenting the longitudinal friction force while the user remains seated.
When we present the longitudinal friction force, we do not have to directly counter the weight of {\color{blue}the} leg in the vertical direction.
Thus, a simple walking display can be realized.
{\color{blue}Also}, we consider the impulse by friction forces in force rendering and {\color{red}provide} a rigorous user study to investigate the effectiveness of the walking sensation display.

\section{Measurement Test of Friction Force}
\subsection{Measurement}
\begin{figure}[t]
  \centering
  \includegraphics[width=1.2in]{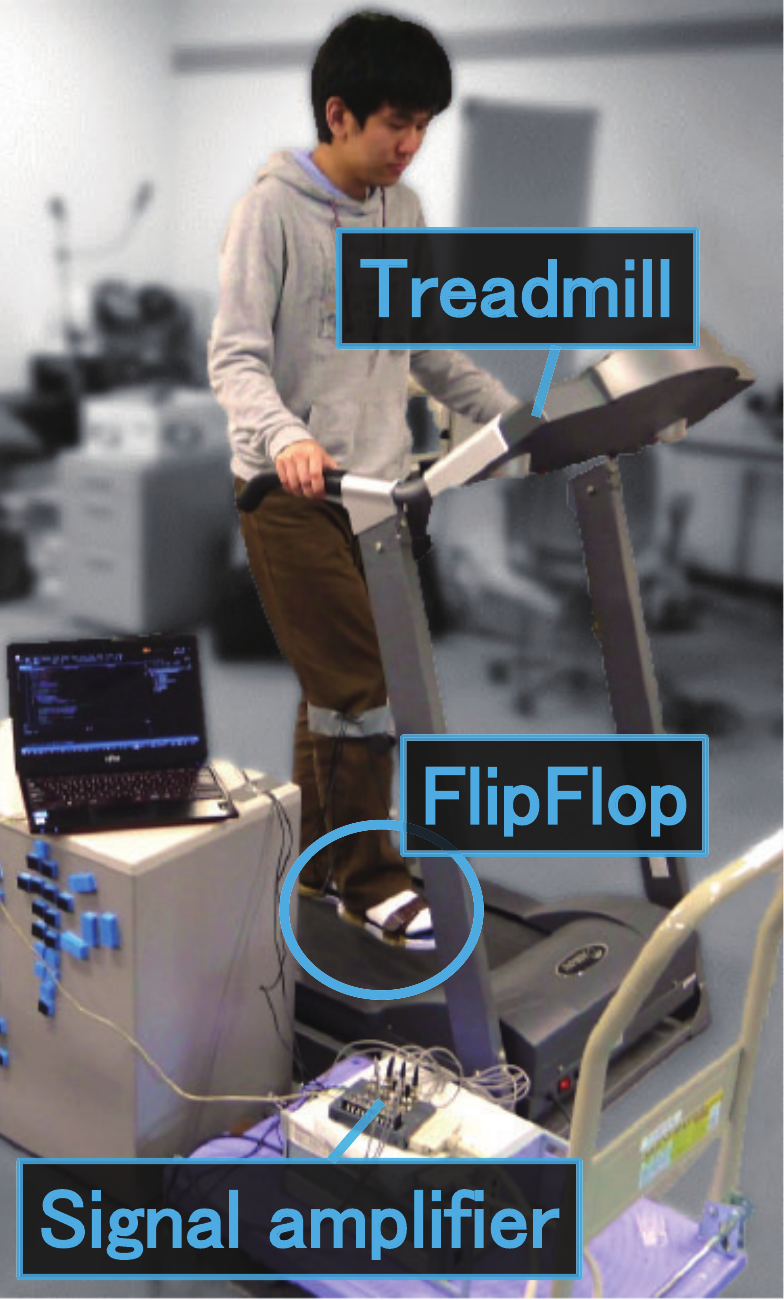}
  \includegraphics[width=2in]{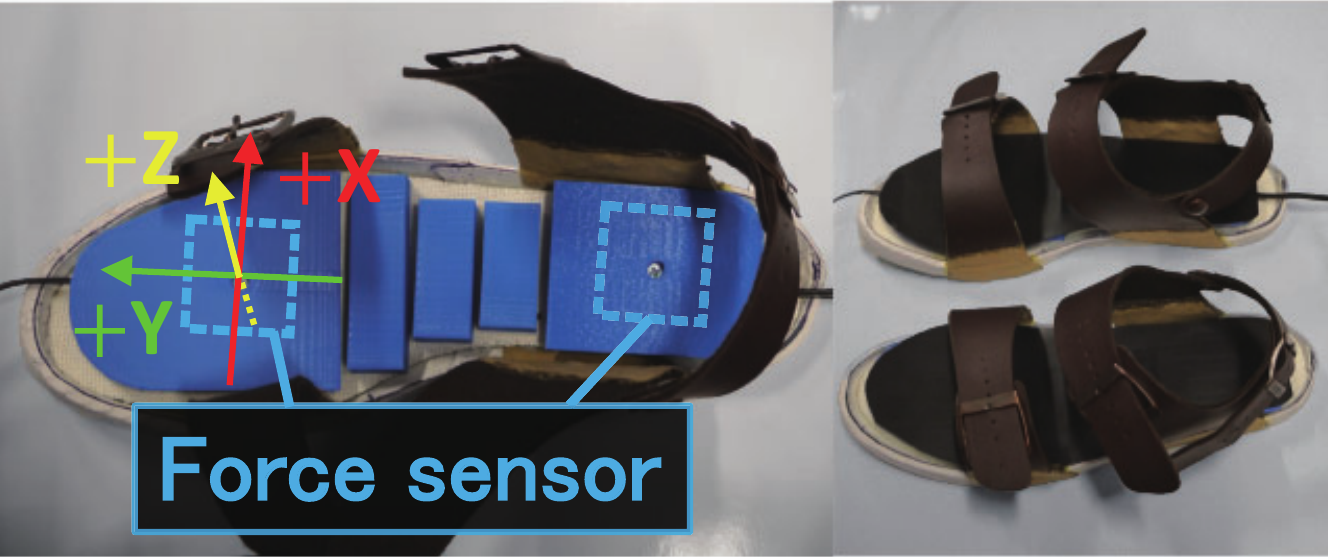} \\
  \hspace{-10mm}(a)\hspace{38mm}(b) \vspace{-3mm}
  \caption{Friction force measurement test when walking (a) experimental situation (b) placement of the force sensors 
  \label{PressureMeasurement}} \vspace{-2mm}
\end{figure}

\begin{figure}[t]
  \centering
  \includegraphics[width=3in]{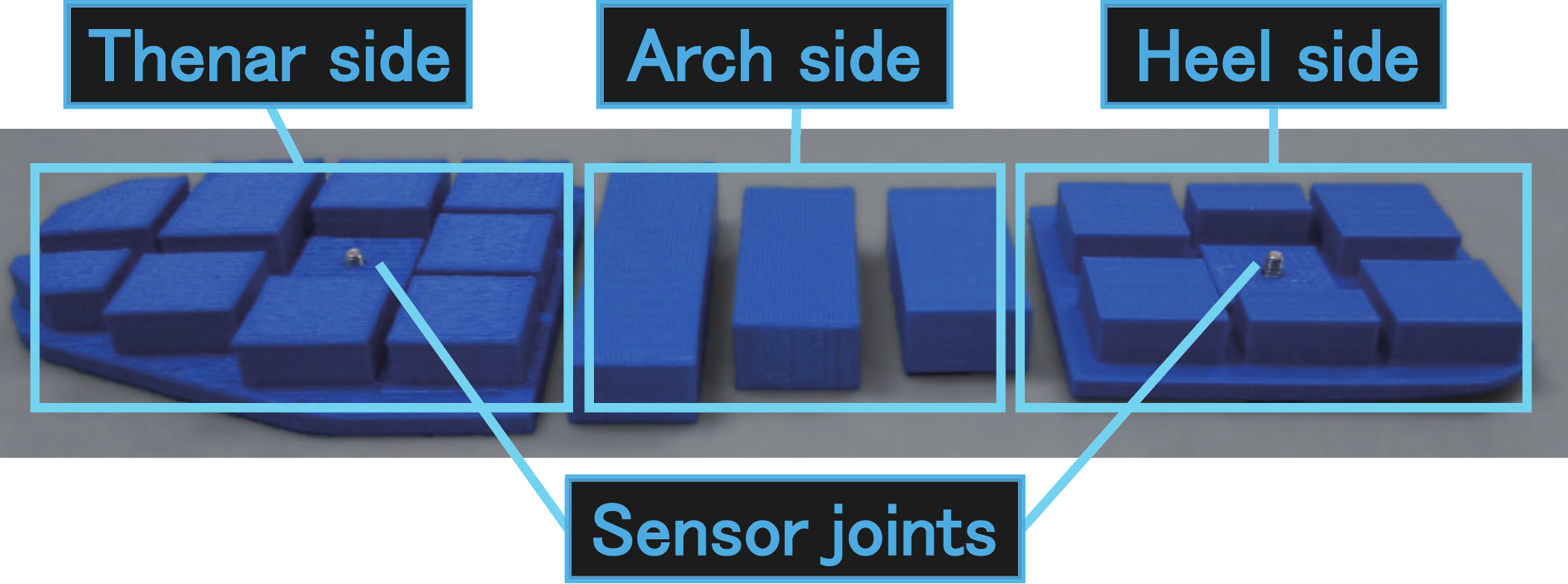} \vspace{-3mm}
  \caption{The spacers for the friction force measurement test \label{Spacer}} \vspace{-5mm}
\end{figure}

To represent the walking sensation by presenting the friction force, 
we {\color{blue}first need} to know the actual friction force applied on the sole of the foot during walking.
{\color{red} 
Ground reaction force, which has been measured classically~\cite{Ground,SShoes}, is the counteracted force against the action forces exerted by the sole of the shoe on the ground.
The ``pressure"} applied on the sole of the foot during walking {\color{blue}has been previously examined~\cite{Plantar,Loading}}.
However,
{\color{blue}to our best knowledge, no study has} measured the {\color{red}friction force on the sole of the foot} during walking.
{\color{blue}Therefore}, we {\color{blue}first made measurements of this force.} 
Fig.~\ref{PressureMeasurement}~(a) shows the experimental situation.
As in Fig.~\ref{PressureMeasurement}~(b), we used {\color{blue}a {\color{red}sandal} with flat insoles, and placed
two three-axis force sensors (USL06-H5-50N, by Tec Gihan Co., Ltd.); one where 
the participants put their thenar of the foot 
and the other where 
they put their heel.} 
{\color{blue}To} prevent the {\color{blue}user's} weight from concentrating on the force sensor, we placed
spacers that cover the sensors (Fig.~\ref{Spacer}) and flattened the part on which 
the participants put their foot. We placed the spacer carefully 
{\color{blue}so as} not to impair the flexibility of {\color{blue}the} {\color{red}sandal} 
and placed the rubber sheet (GS-08, by WAKI SANGYO Co., Ltd.) on the spacer 
to avoid {\color{blue}the user's foot from slipping}.
The Z-axis{\color{red}'} {\color{blue}negative} direction of the force sensor 
corresponds to the direction of gravity, and the Y-axis{\color{red}'} {\color{blue}positive} direction corresponds to the forward 
direction of walking. We used a treadmill to measure the friction force when walking at {\color{blue}1.0, 
2.5}, and 4.0 km/h. 
{\color{red}4.0 km/h is the speed that is generally self-selected when walking without any constraints on a flat surface, and is also used in the study of displaying footstep sounds~\cite{FootstepSounds}. Other speeds (i.e. 1.0 km/h and 2.5 km/h) were chosen as we determined the number of speed conditions was three due to the limited experimental period, and the constant speed difference between conditions was set to 1.5 km/h. }
The treadmill permanently moves the belt of {\color{blue}the} walking space backward{\color{blue}. 
Thus,} the forward friction force will be measured {\color{blue}as} smaller than the force when walking over the ground, 
and the backward friction force will be measured {\color{blue}as larger}. However, there is a correlation between 
the force 
on a treadmill and the force 
during walking over the ground~\cite{Treadmil}. Thus, we prioritized the 
alignment of walking speed. We explain the correction method of this error in section 4.3.

Four participants (average age {\color{blue}$23.3\pm0.5$ years}) joined this measurement. 
They {\color{blue}each} walked 30 steps (each foot landed 15 times) for each walking speed, and we 
extracted the {\color{red}middle} ten steps (from the 4th step to the 13th step) for the analysis {\color{red}because we focus on the steady walking of the central part of the walk}.
To get used to walking with {\color{blue}a} {\color{red}sandal} on the treadmill, 
{\color{blue}the participants practiced for about 10 s before the measurement of each speed.
The participants also took a 30 s break between the measurement of each speed.}

\begin{figure}[t]
  \centering
  \includegraphics[width=1.6in]{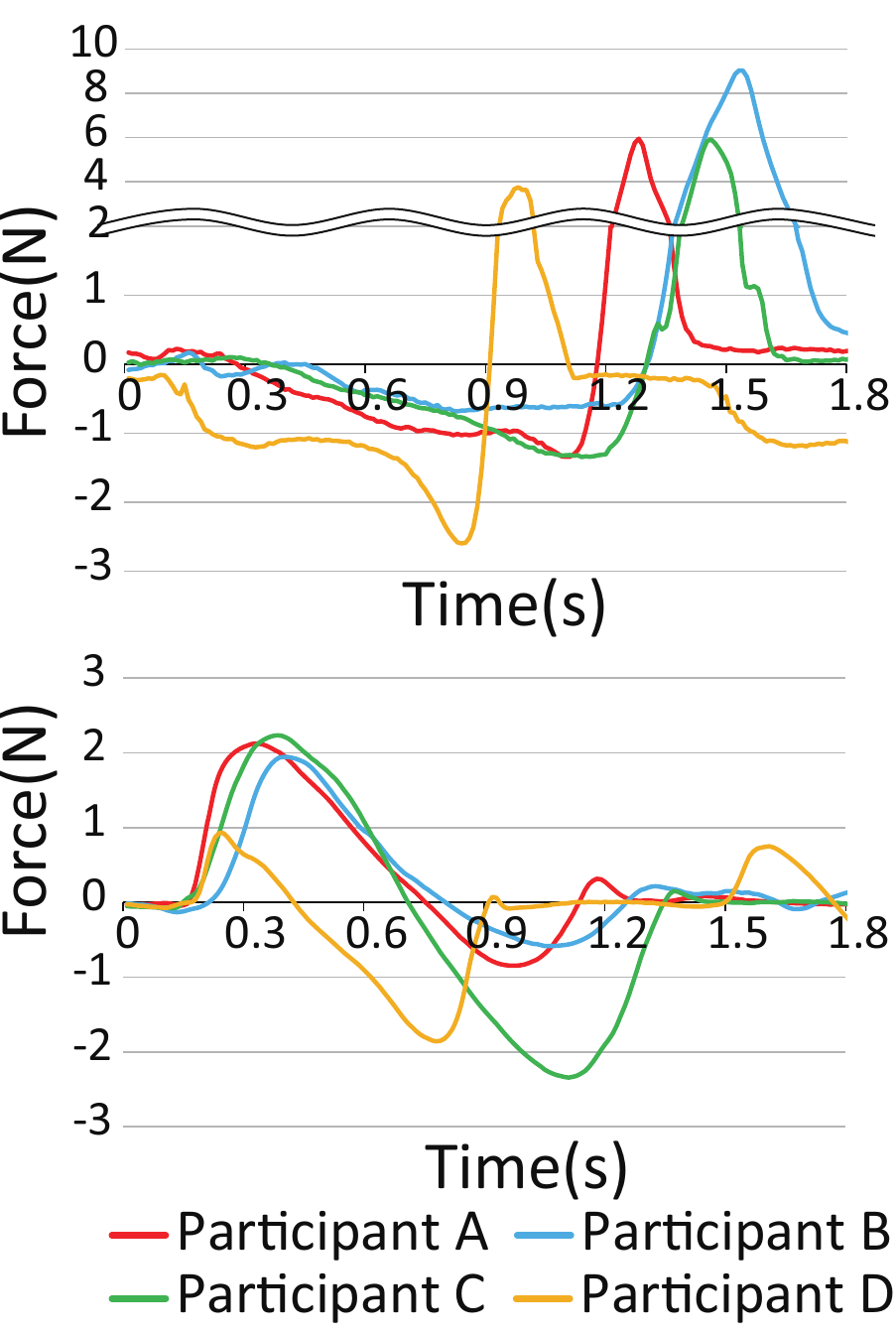}
  \includegraphics[width=1.6in]{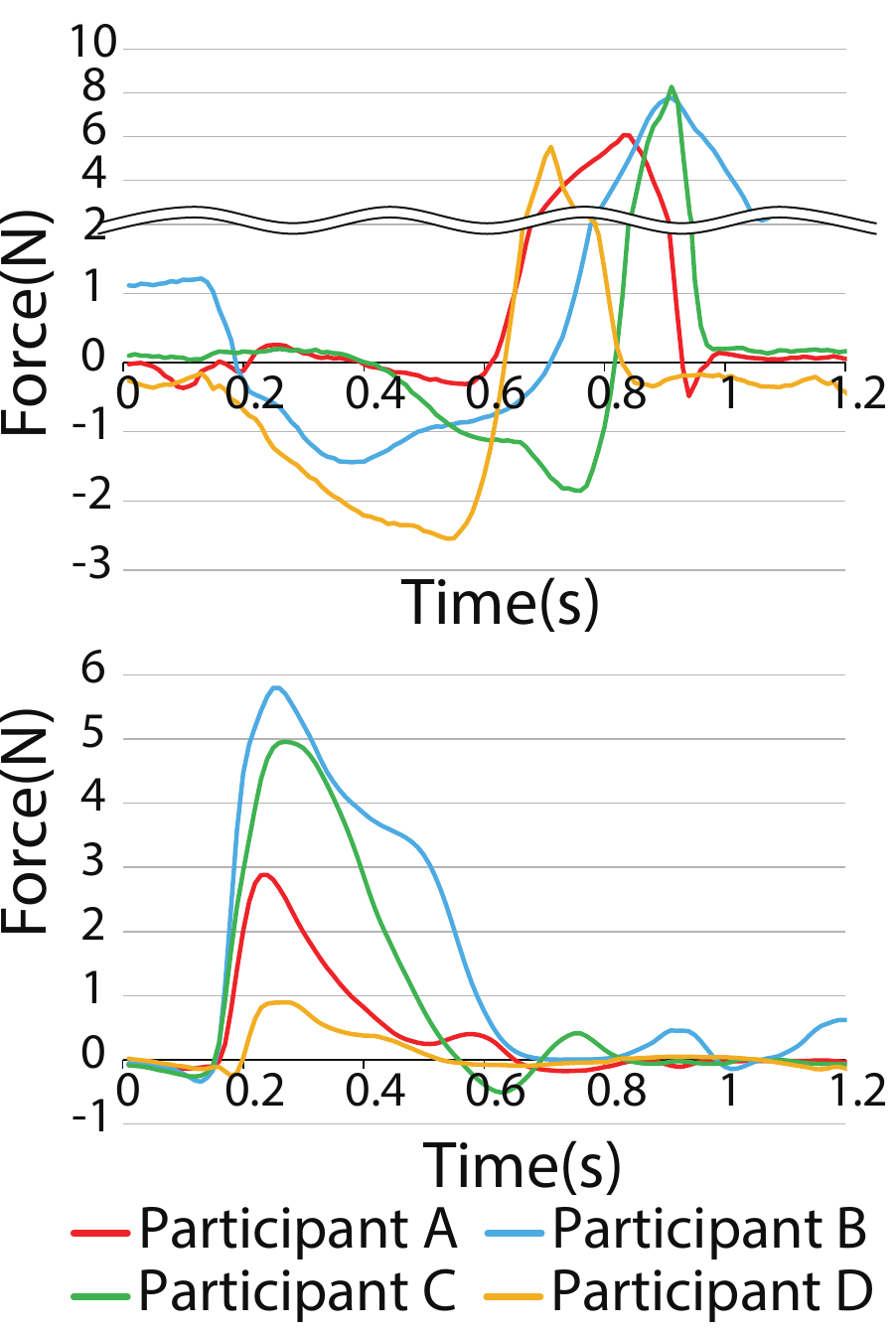} \\
  \hspace{9mm}(1.0 km/h)\hspace{29mm}(2.5 km/h) \\
  \vspace{2mm}  
  \includegraphics[width=1.6in]{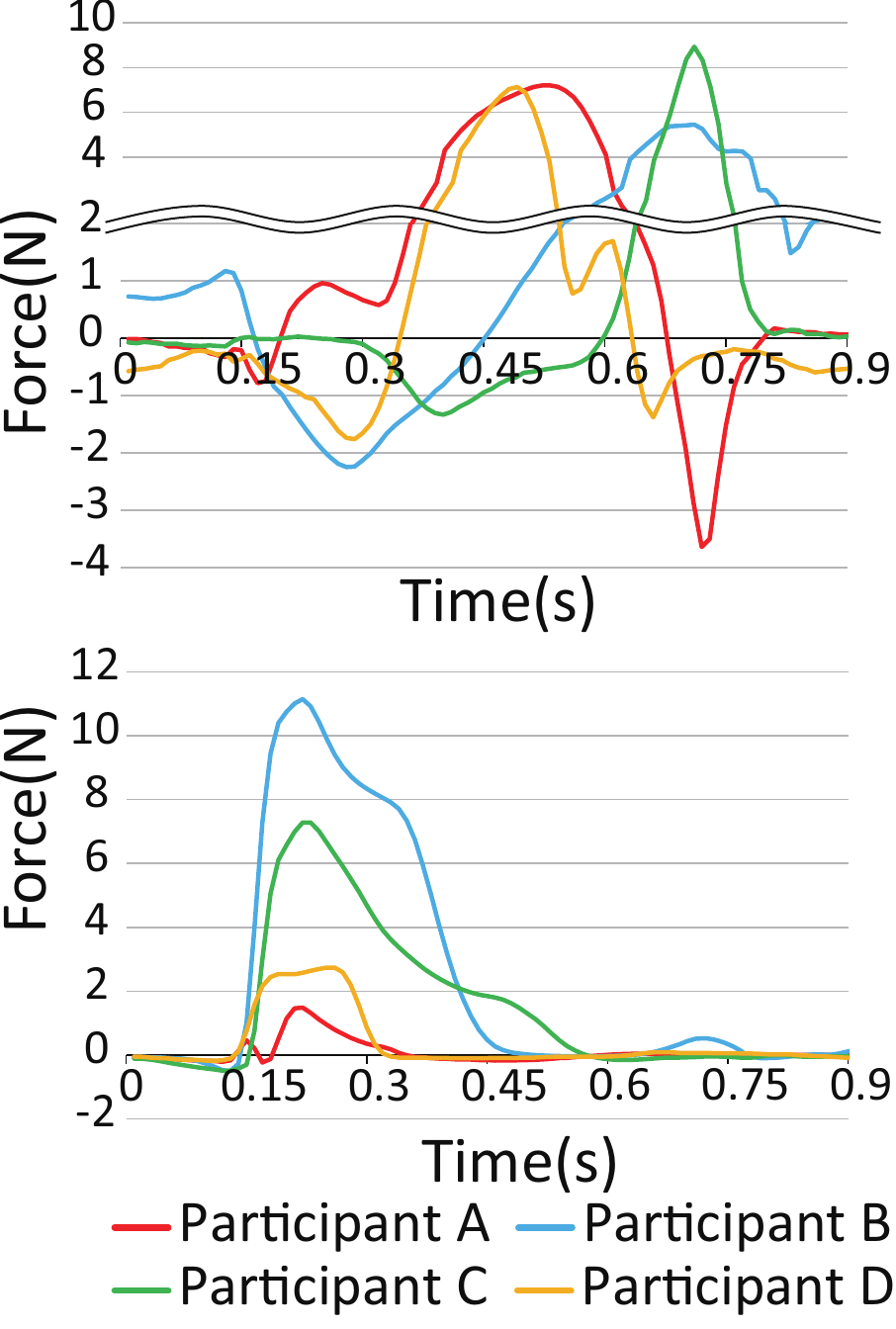} \\
  \hspace{5mm}(4.0 km/h)
  \vspace{-3mm}
  \caption{The result of the friction force measurement test when walking (Upper: thenar side, Lower: heel side) \label{FootWave}} \vspace{-5mm}
\end{figure}

Fig.~\ref{FootWave} shows the measured friction force of each walking speed.
In these graphs, we show the averaged waves of ten steps in the Y-axis direction.
As described below, extremely strong load was applied on the thenar side at the end of each step.
Thus, the scale of the values stronger than 2 N in {\color{blue}the} thenar side graphs was set differently.
The actual force applied on the sole corresponds to the reaction force of the measured force by the sensors. Thus, the friction forces applied on the sole {\color{blue}were}
obtained by reversing the sign of the forces in Fig.~\ref{FootWave}.
In other words, the {\color{blue}positive} values in this figure {\color{blue}show} the backward friction forces, 
and the {\color{blue}negative} values in this figure {\color{blue}show} the forward friction forces.

\subsection{{\color{red}Discussion}}
\begin{figure}[t]
  \centering
  \includegraphics[width=3.4in]{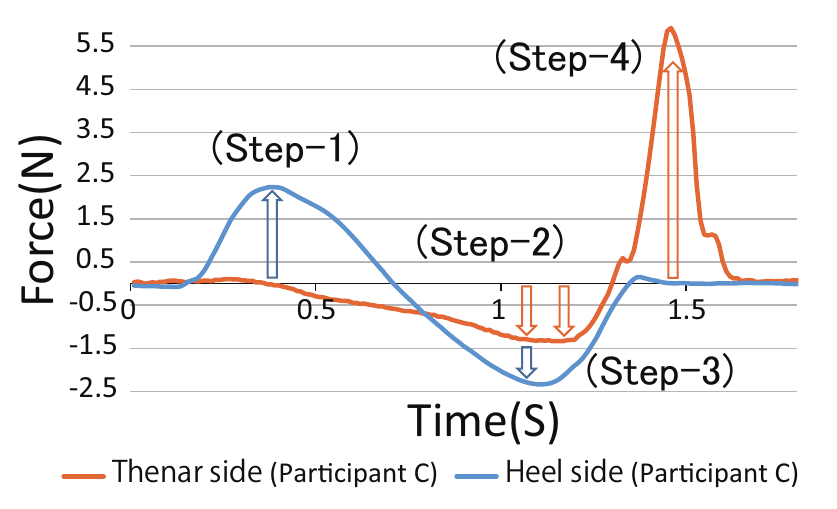} \vspace{-5mm}
  \caption{A certain participant's result of the measurement test
  \label{FootWaveOne}} \vspace{-4mm}
\end{figure}

\begin{figure}[t]
  \centering
  \includegraphics[width=3.4in]{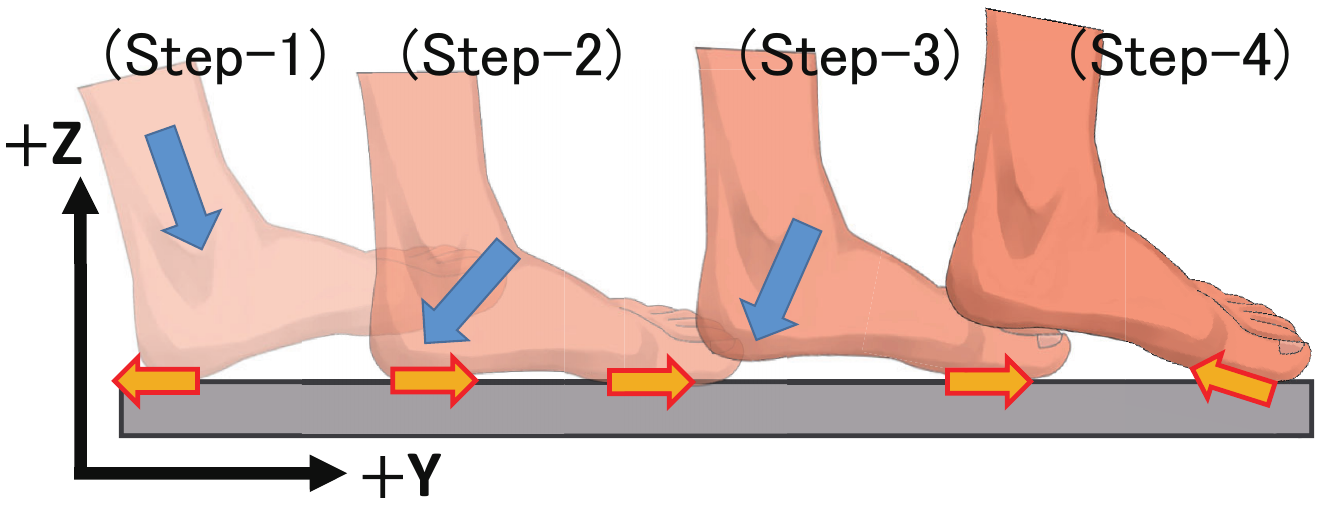} \vspace{-5mm}
  \caption{The forces inferred to be applied on the sole of the foot in one step (Orange arrows: friction force on the sole of the foot, Blue arrows: the force inferred that the human applies to the foot)  \label{FootMove}} \vspace{-2mm}
\end{figure}

As a result of the measurement test, we could confirm the regularity in the transition of the friction force 
during walking. As a representative example, {\color{blue}in Fig.~\ref{FootWaveOne},} we show the {\color{blue}values} at 1.0 km/h of a certain participant. 
Fig.~\ref{FootMove} shows the outline of the applied friction force inferred from the measurement. 

At the beginning of the {\color{blue}first} step, (Step-1){\color{blue},} a backward force is applied on the heel side, and then {\color{red}(}Step-2) 
forward force is applied on both the heel side and thenar side. 
{\color{blue}Then (Step-3),} the forward force on the heel side becomes small. 
Finally {\color{blue} (Step-4), a} large backward force is suddenly applied on the thenar side.
The friction force in (Step-1) is inferred as the reaction force against the force
{\color{blue}used} to put the foot forward. This force is inferred to act as a brake of the body
and {\color{blue}to} stabilize the walking motion. The friction forces in (Step-2) and (Step-3) are
inferred as the reaction force against the backward force applied when {\color{blue}kicking} the ground. 
This reaction force acts as a driving force to move the body forward. As the weight
shifts from the heel side to the thenar side, {\color{red}the heel is lifted.} 
Thus, in (Step-3), the force {\color{blue}is} applied only on the thenar side. As the angle between the foot
and the plate changes, the weight of the body is applied not only in a vertical direction (Z-axis) but also
in a horizontal direction (Y-axis). This is considered as the cause of the sudden change {\color{blue}in force} (Step-4). 
Thus, the force is applied by the longitudinal friction force only in (Step-1), (Step-2), and (Step-3).
When we increased the speed of the treadmill during the measurement, the forward force applied on both the heel and thenar sides like (Step-2) {\color{blue}does} not appear
because of {\color{blue}a reduction in the time from when the heel side is grounded to the foot leaving the ground. Except for
(Step-2), similarly} shaped forces were applied regardless of the speeds and participants.


\section{Development of a Friction-based Walking Sensation Display for a Seated User}
\subsection{Hardware Design}
\begin{figure}[t]
  \centering
  \includegraphics[width=1.7in]{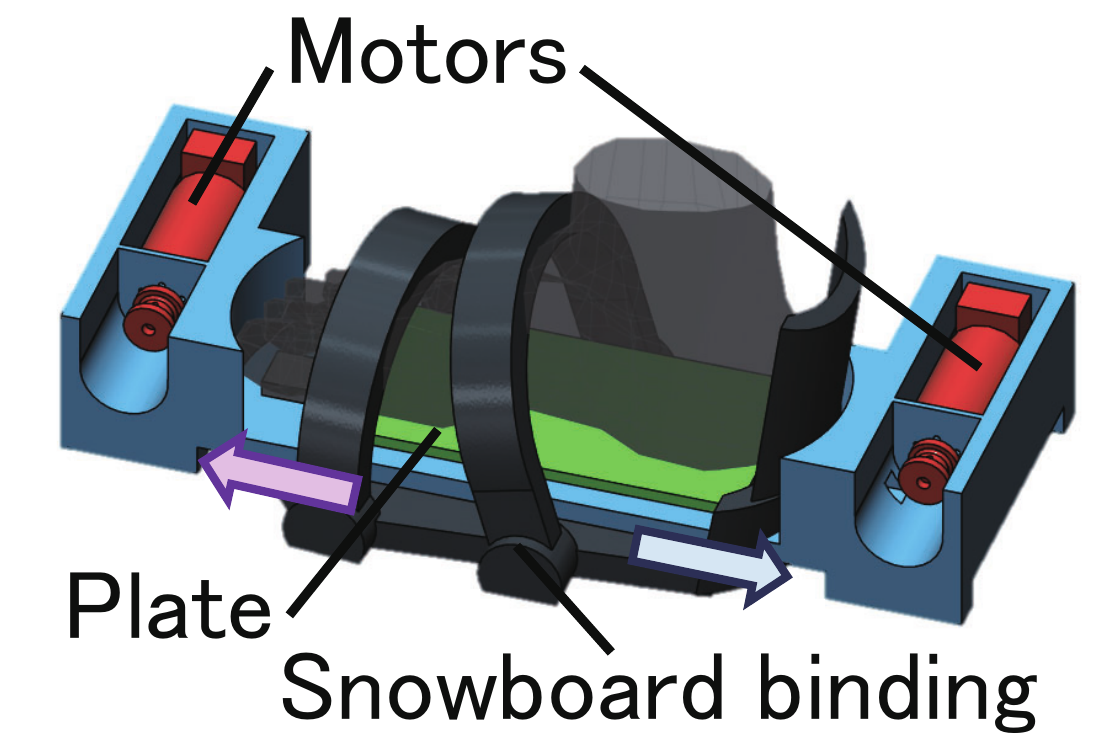}
  \hspace{-1.5mm}
  \includegraphics[width=1.6in]{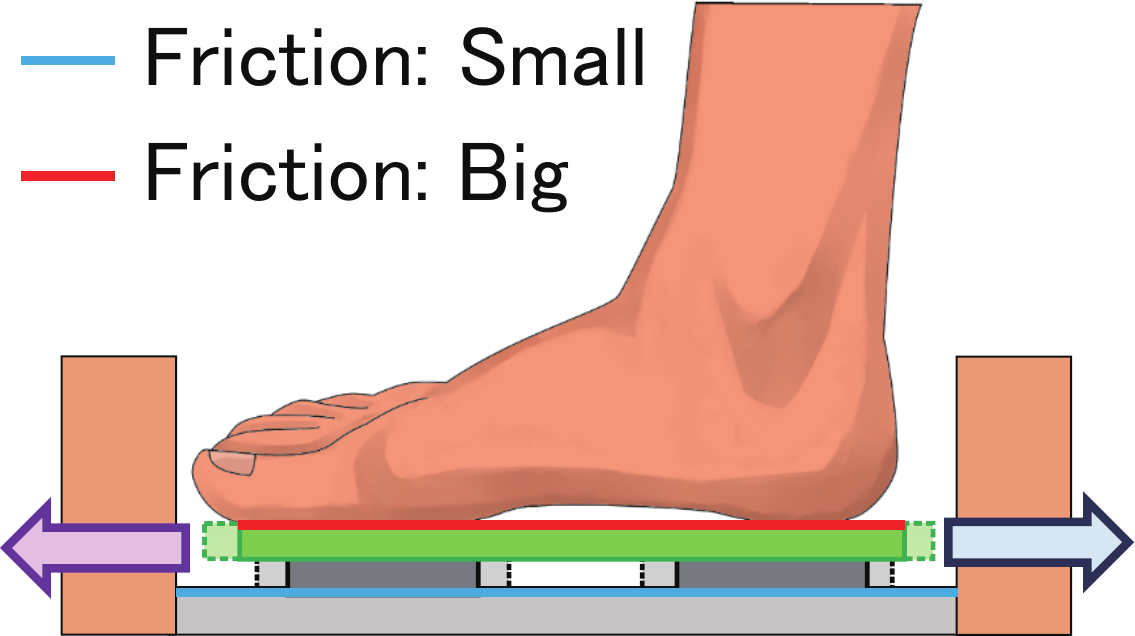}\\
  \hspace{4mm}(a)\hspace{38mm}(b) \vspace{-3mm}
  \caption{(a) The outline of linear rail mechanism (b) The structure of the proposed device. \label{HapStep}} \vspace{-5mm}
\end{figure}

{\color{blue}Because we detected} regularity in the transition of the friction force during walking, we {\color{blue}attempted} to represent the walking sensation by mimicking the measured friction force.
Specifically, we {\color{blue}attempted} to represent the friction force in (Step-1), (Step-2), and (Step-3){\color{blue}, as}
{\color{red}described} in the previous section. 
We do not represent the friction force in (Step-4) because we {\color{blue}believe} this force is caused by {\color{blue}body weight load}. {\color{red}The reaction force to support the body weight itself will be important for representing the walking sensation. However, in our method, we do not present the upward force. Thus, 
if we present the reaction force to the 
body weight only when the foot leaves ground, this would likely confuse the user.} As mentioned above, the transitions of the friction force {\color{blue}differ} between 
the thenar side and the heel side. However, 
{\color{blue}to simplify the device and attempt to represent the walking sensation using the 1-DOF simple friction force display, 
in this study, we regarded the entire sole as the one presentation area of the friction force.}
{\color{red} We first present the friction force forward and then backward.}
We developed a haptic device to realize the 1-DOF friction force. 
Fig.~\ref{HapStep} (a) shows the entire outline of the proposed device.
As shown in Fig.~\ref{HapStep} (b), {\color{blue}the user's foot is placed on a} plate on a linear rail
(SRS9XN, by THK Co. Ltd). By winding the wire (HARDCORE X8 \#6.0, by DUEL CO. Inc) 
attached to the plate with a motor (RE40 148877, by Maxon Motor AG),
a friction force is applied to the sole.
The friction between the linear rail and the movable part is extremely small (dynamic friction coefficient $\mu \leq 0.003$, which is much smaller than the coefficient {\color{red} of the friction that works} between typical ice and iron, 0.027). 
{\color{blue}Thus,} this device can 
present the friction force without the influence of the leg's weight, and 
{\color{blue}(because the resistance force is small) this can be done rapidly.}
We controlled the motors {\color{blue}using a} microcomputer (Arduino Uno, by Arduino S.R.L.).
By sending the signal from the microcomputer to the motor driver (MD10C, by Cytron Technologies{\color{red})} 
connected to the 10~V power supply, {\color{blue}a} voltage between 0 and 10~V is applied to the motors.
To fix the user's leg to the frame, we {\color{blue}used a} snowboard binding (ZUMA BLACK S/M ZUMA board binding, 
Swallow Ski Co. Ltd).

\subsection{Performance Test}
\subsubsection{{\color{red}Experimental Setup}}
\begin{figure}[t]
  \centering
  \includegraphics[width=1.6in]{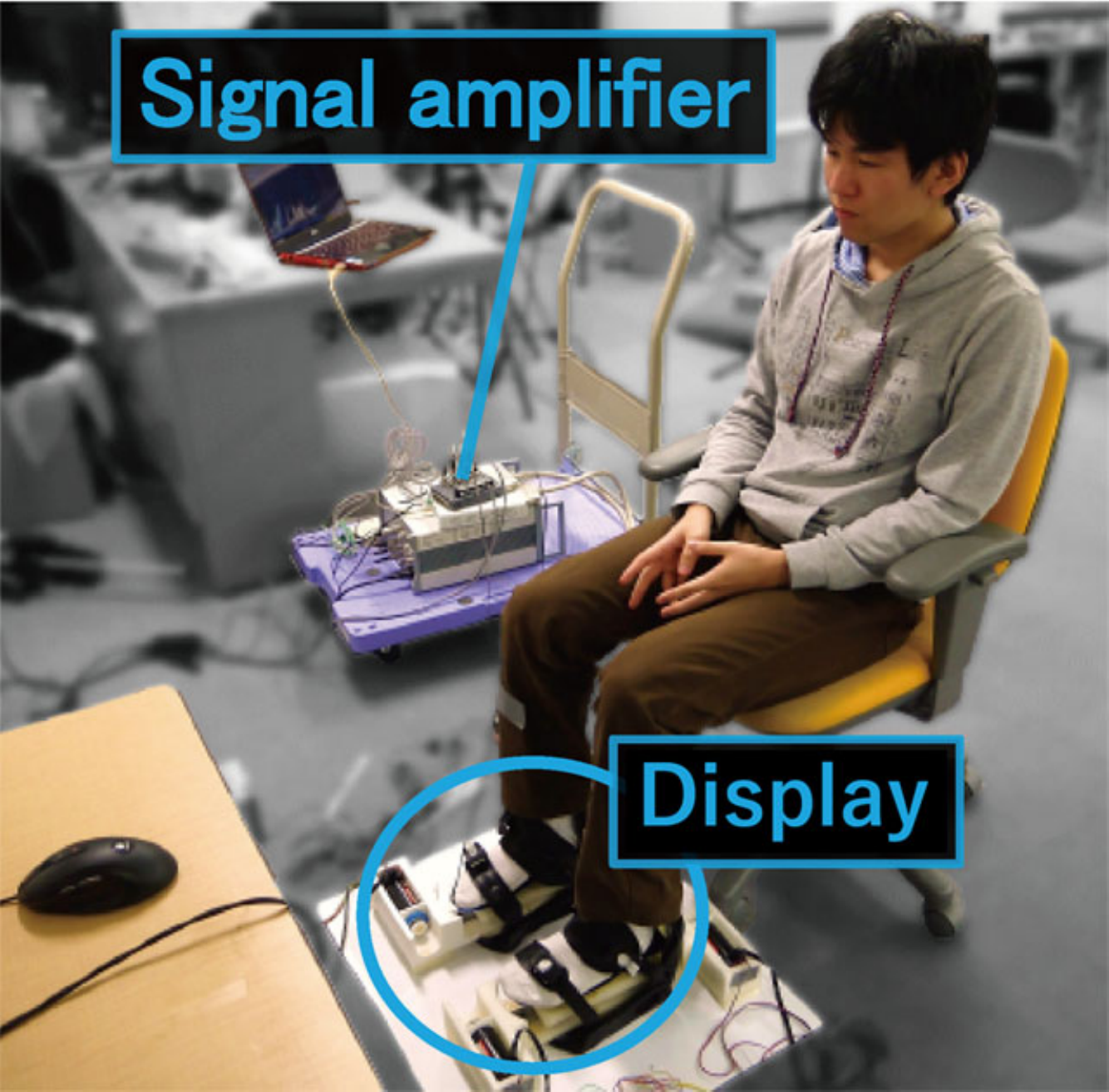}
  \includegraphics[width=1.6in]{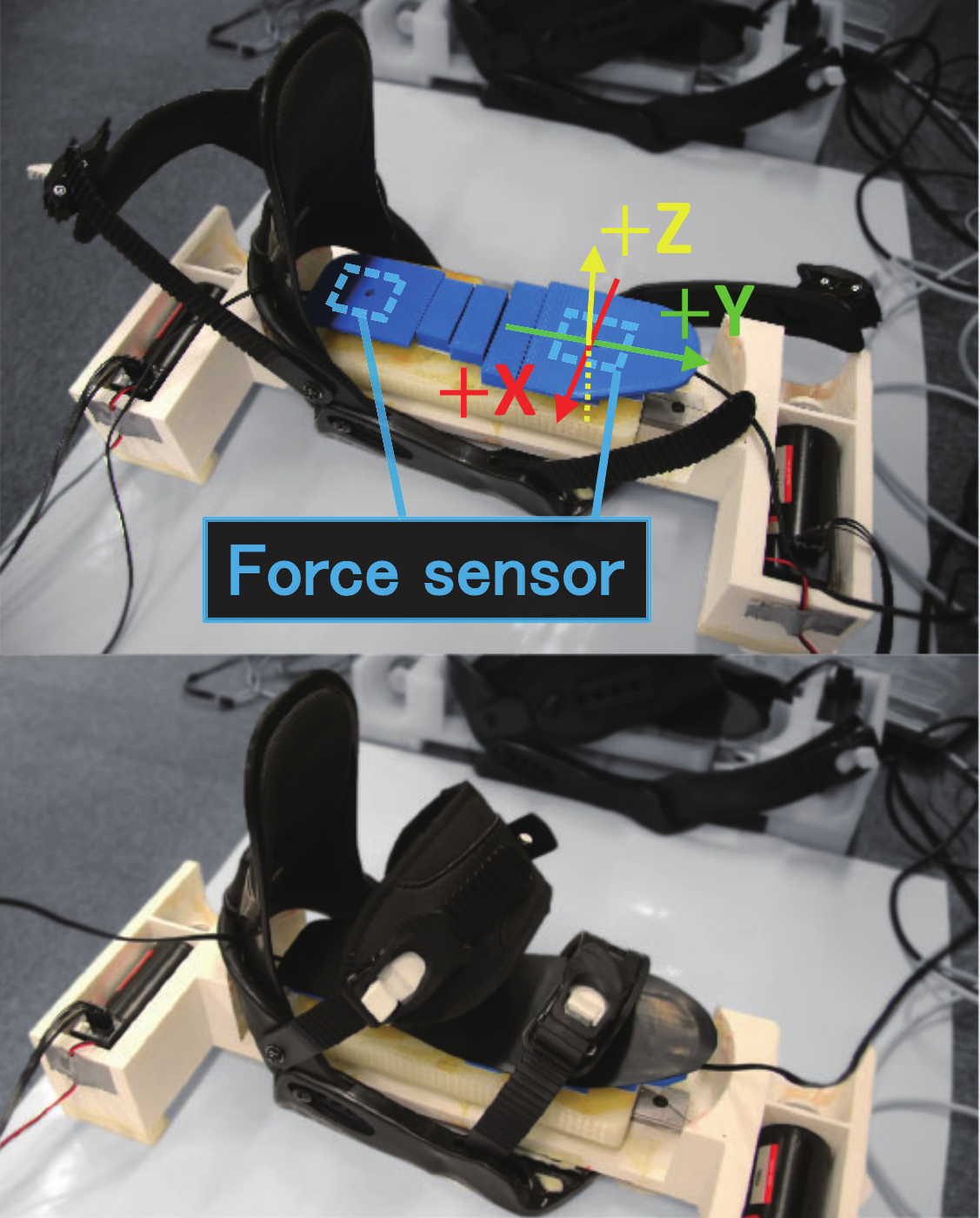}\\
  \hspace{2mm}(a)\hspace{40mm}(b) \vspace{-3mm}
  \caption{Friction force measurement test of the proposed display (a) experimental situation (b) placement of the force sensors \label{PressureMeasurementHapStep}} \vspace{-2mm}
\end{figure}

\begin{figure}[t]
  \centering
  \includegraphics[width=1.8in]{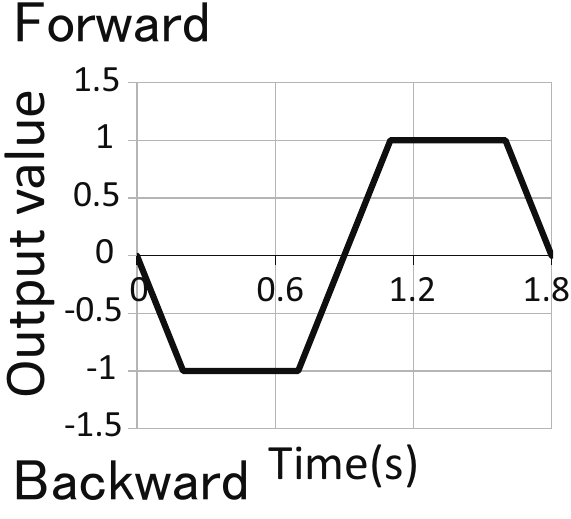} \vspace{-3.5mm}
  \caption{The transition of the output signal in the friction force measurement test of the proposed display
  \label{InputWaveSlow}} \vspace{-2mm}
\end{figure}

To investigate the friction force that the proposed display can present{\color{blue}, as well as} the responsiveness of the device, we measured 
the friction force {\color{blue}generated by the device using the same sensor described in Section 3}.
Fig.~\ref{PressureMeasurementHapStep} shows the experimental situation and the placement of the force sensors.
{\color{blue}As in Section 3}, we placed the sensors on the thenar side and the heel side
and covered them {\color{blue}with a} rubber sheet.
{\color{blue}Because we planned} to represent the walking sensation by presenting of friction force 
in the order of ``backward" then ``forward", we measured the friction force by applying the load in this order. 
Fig.~\ref{InputWaveSlow} shows the transition of the output signal in this measurement.
The {\color{blue}positive} value shows the signal to the motor towing the plate forward (forward motor) while the {\color{blue}negative} value shows the signal to the motor towing backward (backward motor).
At first, we {\color{blue}increased} the output value to the backward motor linearly to reach the maximum value in 0.2~s.
Next, after {\color{blue}maintaining} the maximum value for 0.5 s, we decrease the value linearly to {\color{blue}zero} in 0.2~s.
After this procedure, we input the same signal to the forward motor.

The rise and the fall times (0.2~s) were determined to ensure the output stabilities.
{\color{blue}Also}, the presentation time of the maximum load (0.5~s) was determined to ensure that 
we {\color{blue}could collect sufficient} data. We controlled the load of the motors by the PWM (Pulse Width Modulation) signals from the microcomputer (Arduino Uno)
and used the three PWM outputs of {\color{blue}the} Arduino Uno (95, 175, and 255) as the maximum output signals (corresponding to the 
output value 1.0 in Fig.~\ref{InputWaveSlow}). 
These PWM signals approximately {\color{blue}correspond to the duty rates (0.37, 0.69, and 1, respectively). 
When the duty rate is 1, 10~V is continuously applied to the motor}.
The minimum signal {\color{blue}(95)} was determined to ensure that the proposed device can move the skin of the sole, regardless of the participant. 
The medium signal (175) was the average value of the minimum (95) 
and maximum (255) values. The same four participants {\color{blue}of Section 3 joined here.} 
We measured the friction force ten times for each signal pattern (10 times $\times$ 3 signal patterns $=$ 30 trials).

\subsubsection{{\color{red}Results and Discussion}}
\begin{figure}[t]
  \centering
  \includegraphics[width=3.5in]{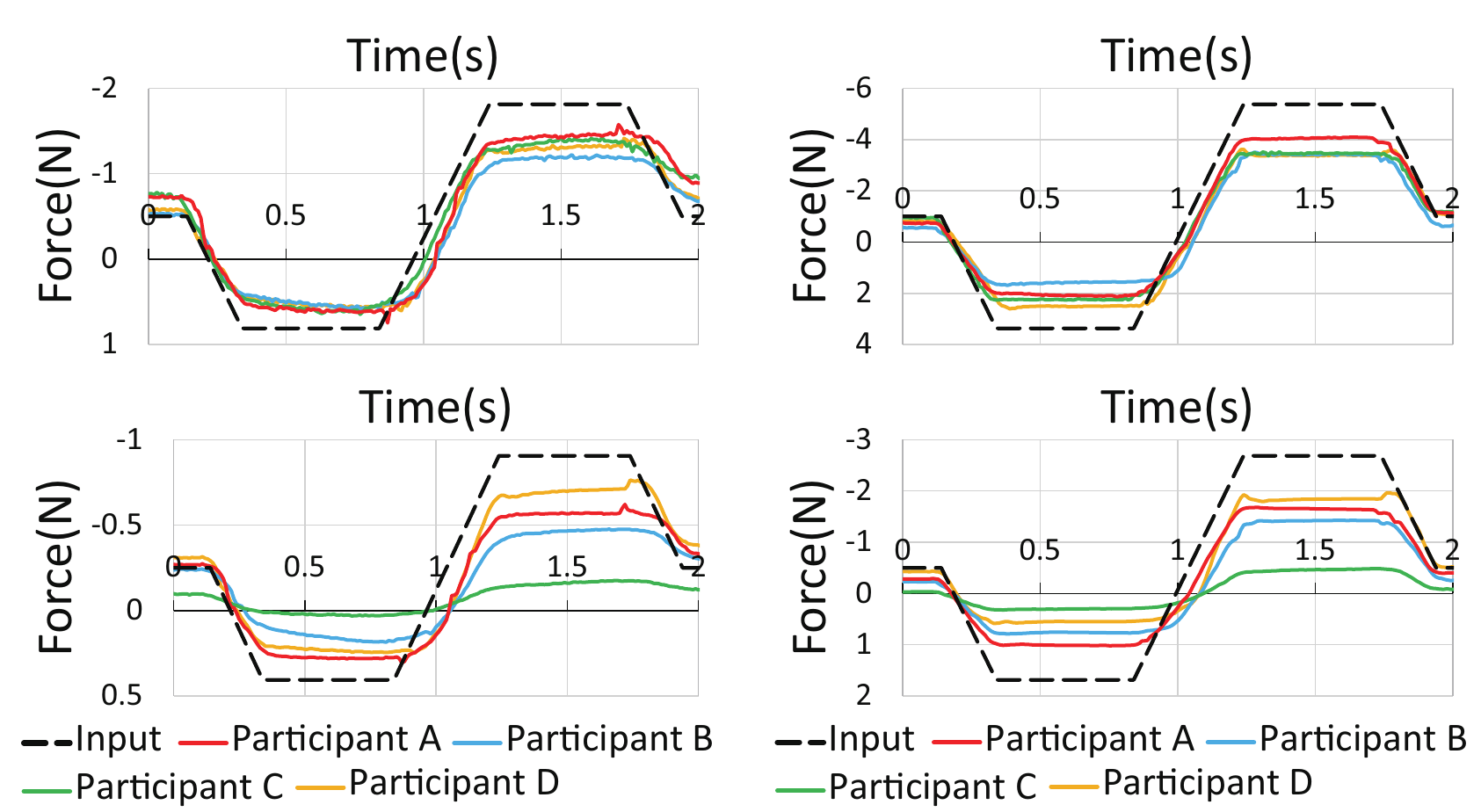} \vspace{-2mm}
  \hspace{4mm}(95/255)\hspace{35mm}(255/255) \vspace{-0mm}
  \caption{The result of the friction force measurement test of the proposed display (Upper: thenar side, Lower: heel side) \label{PressureWaveSlow}} \vspace{-2.75mm}
\end{figure}

Fig.~\ref{PressureWaveSlow} shows the averaged result of each signal pattern.
We show the averaged point of ten measurements.
As representative examples, we show the results when the maximum signals are 95 and 255.
From this figure, we found that {\color{blue}although} the variance of the friction force on the thenar side was 
small, the friction force on the heel side varied depending on the {\color{blue}participant}.
{\color{blue}Because of the structure of the proposed display, 
there were some differences in the fixation of the foot depending on the participant}. 
The structure around the heel and ankle is surrounded by the front and back fixtures 
(Fig.~\ref{HapStep} (b)). Thus the fixation was influenced by the shape of the foot.
On the other hand, the thenar side is fixed only by the single belt.
We believe that this made the influence of the {\color{blue}foot's} shape small, and also {\color{blue}reduced the variance of the friction forces}.

\begin{figure}[t]
  \centering
  \includegraphics[width=2.9in]{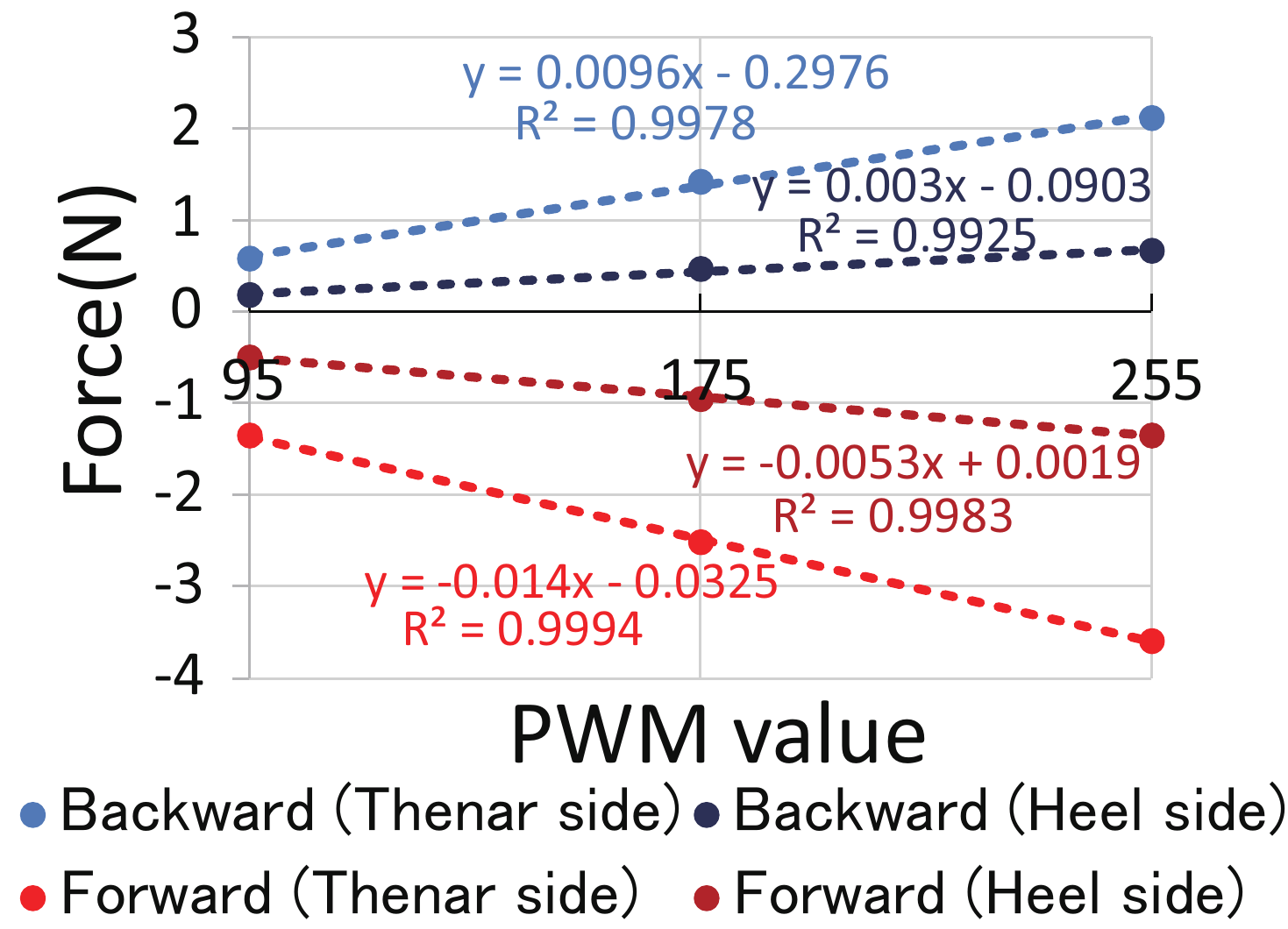} \vspace{-3mm}
  \caption{The proportional relations between the averaged peak friction forces and the PWM output signals \label{LinearRelation1}} \vspace{-3mm}
\end{figure}

\begin{figure}[t]
  \centering
  \includegraphics[width=2.8in]{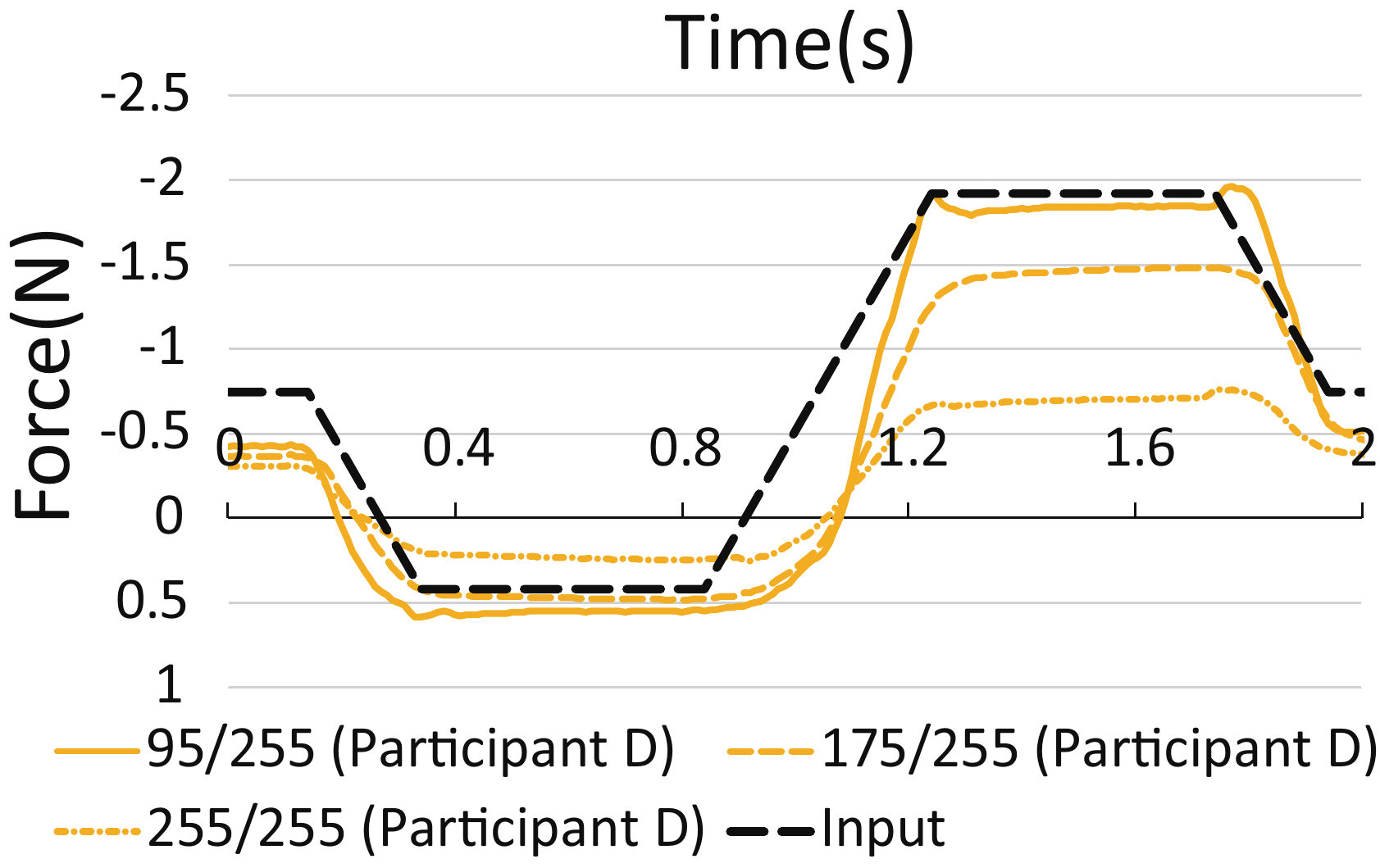} \vspace{-3mm}
  \caption{The measured friction force on the heel side that did not change very 
  much even if the duty rate was changed \label{LinearRelation2}} \vspace{-2mm}
\end{figure}

Fig.~\ref{LinearRelation1} shows the proportional relation between the backward/forward friction 
forces and the output signals. We calculated this relation based on the averaged peak loads of all participants.
In this figure, we show the PWM signal value while its divided value by 255 corresponds to the duty rate. 
We {\color{blue}found} that the magnitudes of the forces applied to the thenar side 
and the heel side are linearly proportional to the duty rate. In fact, as shown in Fig.~\ref{LinearRelation1},
all of the coefficients of determination were greater than 0.99.
On the other hand, there was {\color{blue}one} participant {\color{blue}for whom the} measured backward friction force on the heel side 
did not change {\color{blue}significantly,} even if the duty rate was changed (Fig.~\ref{LinearRelation2}). 
Due to the structure of the device, the heel side sometimes {\color{blue}rises} a little off the ground when the plate moves backward, 
and the foot is pressed against the fixture. 
This might have caused instability {\color{blue}for} the force 
on the heel side. Considering this problem{\color{blue}, as well as} the problem of the variance {\color{blue}(as mentioned above),} 
we {\color{blue}feel that} it is necessary to improve the structure of the device near the heel side.
However, the floating distance 
was small, and it did not {\color{blue}greatly disturb} 
the friction force.
{\color{blue}Also, although} the variation range was small, we could present {\color{blue}a} friction force proportional 
to the duty rate, even for the participant whose heel side {\color{blue}lifted} off the ground. 
Thus, we found that the proposed device can present the friction force 
proportional to the output signal.

{\color{blue}Also}, in order not to generate the unnecessary load when the output signal is 0, 
the proposed display adjusted the position of the friction display plate only when it 
greatly deviated from the center.
At the time of the preliminary experiment, we found that the plate can return to the center position 
automatically by the elasticity of the skin. {\color{red}Thus,} we decided not to apply the extra force during 
the transition to the no-load state.
As a result, at the moment the proposed display transitioned to the no-load state, the position of the plate did not return to the neutral position
perfectly. This is the reason that the friction force at the no-load state was not 0.
On the other hand, there were no situations {\color{red}where} the plate {\color{blue}is} gradually shifting {\color{blue}in} some directions{\color{blue}.  
Thus,} we {\color{blue}believe that} the proposed display can present the friction force{\color{blue}s} {\color{red}in succession.} 

\begin{figure}[t]
  \centering
  \includegraphics[width=1.6in]{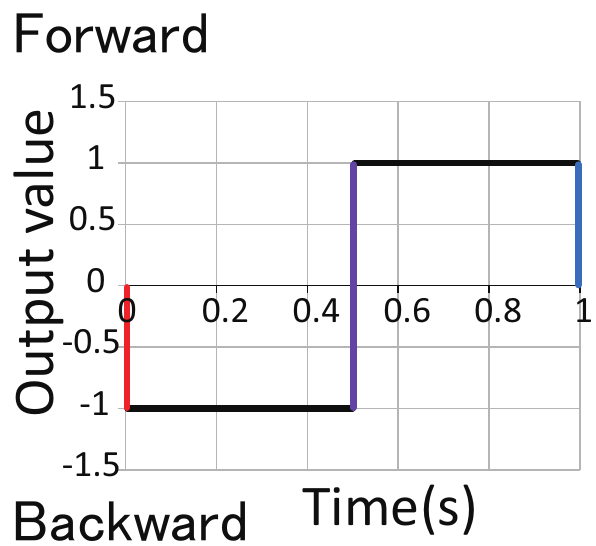}
  \includegraphics[width=1.6in]{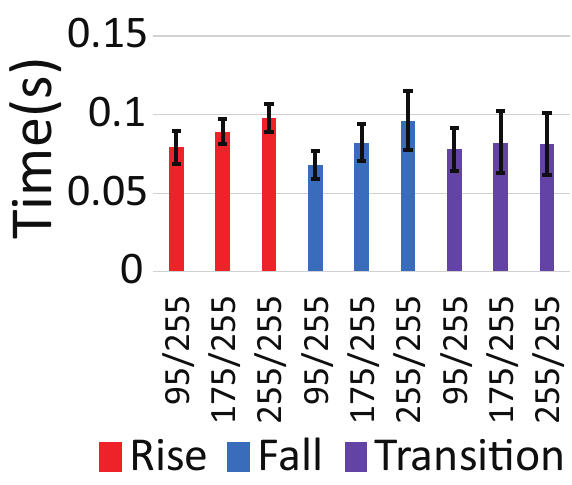}\\
  \hspace{4mm}(a)\hspace{40mm}(b) \vspace{-3mm}
  \caption{The responsiveness investigation of the proposed display (a) the transition of the output signal (b) response time and standard deviation \label{InputFastAndDelay}} \vspace{-1mm}
\end{figure}

Next, to investigate the responsiveness of the proposed display, 
we measured the friction force when the rise and fall times of {\color{blue}the} output signals are 0 s.
As shown in Fig.~\ref{InputFastAndDelay} (a), 
we gave the signal that rapidly changes in 0 s, 
and measured the friction force for the three maximum signals (PWM signal 95, 175, {\color{blue}and} 255).
Fig.~\ref{InputFastAndDelay} (b) shows not only the rise and the fall time{\color{blue}s} of the measured friction force 
but also the transition time from the end of the backward output to the forward peak load.
The time until the value drops for the first time after the output was defined as the rise time.

As shown in Fig.~\ref{InputFastAndDelay}(b), we {\color{blue}found} that the proposed display {\color{blue}responds within 
about 0.1 s,} regardless of the magnitude {\color{blue}or} direction of the friction force.
{\color{blue}Also}, we confirmed that the responsiveness did not deteriorate when the display {\color{blue}subsequently moved the plate  
in the opposite direction.} 
Thus, {\color{blue}the proposed device can rapidly and stably present a friction force}.

\subsection{Calculation of Output Signal}
From previous measurements, we {\color{blue}estimated} the 
friction forces 
during walking 
and the friction forces per duty rate presented by the proposed display.
From these data, 
{\color{blue}to represent the walking sensation,} we calculated the output signal to the proposed display. 

\begin{figure}[!t]
  \centering
  \includegraphics[width=3.2in]{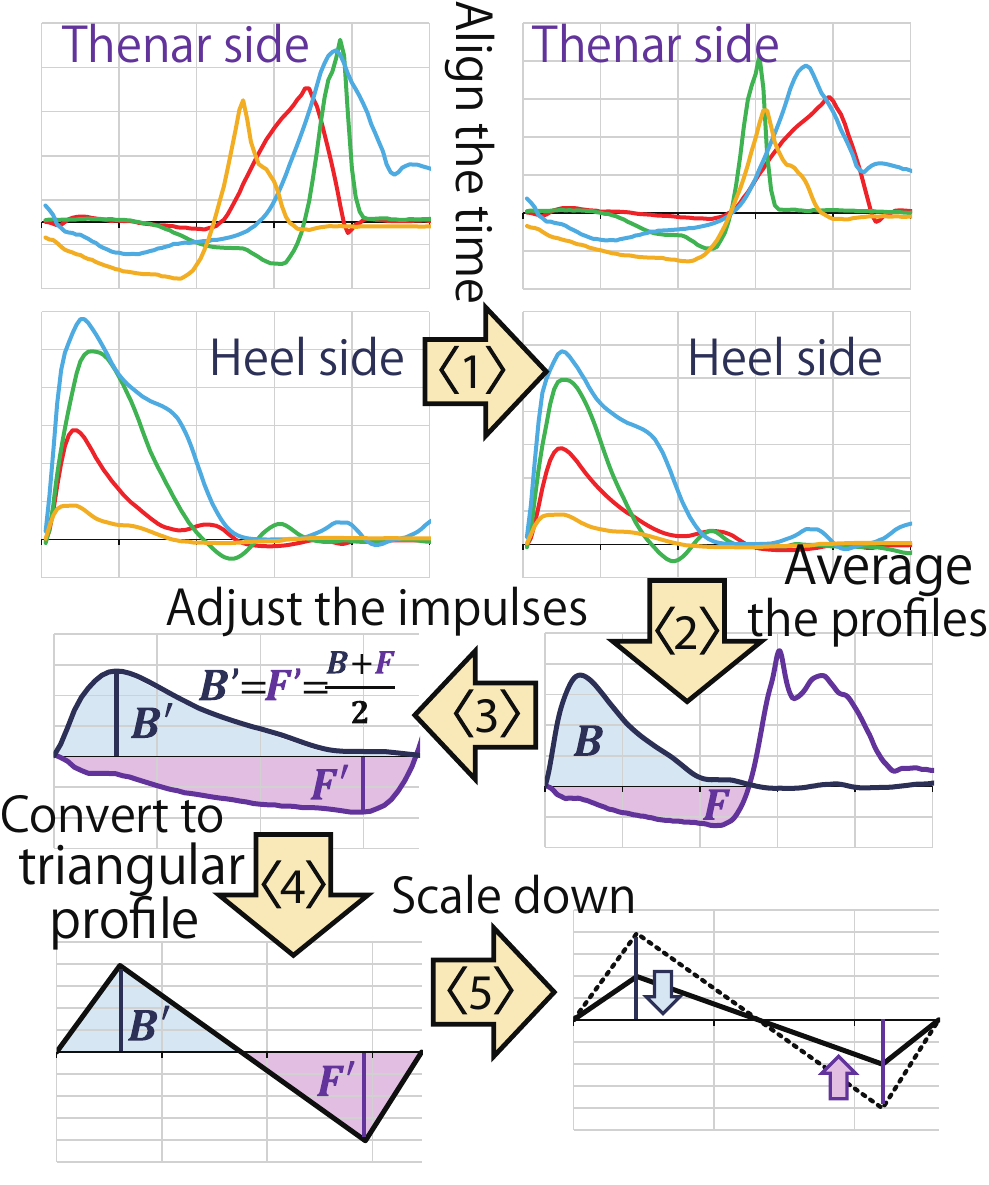} \vspace{-5mm}
  \caption{The procedure to calculate the output signal \label{Calculation}} \vspace{-2mm}
\end{figure}

Fig.~\ref{Calculation} shows 
the outline of the procedure {\color{blue}used} to calculate the output signal from 
the friction force measured during walking at 2.5 km/h.
{\color{blue}Below}, we introduce the specific {\color{blue}steps} of the procedure.

\vspace{-2.5mm}
\begin{description}
\setlength{\itemsep}{-3pt}      
\setlength{\labelsep}{2pt}     
\setlength{\itemindent}{-10pt}   
\setlength{\parskip}{8.5pt}      
\item[$\left< 1 \right>$] Align the time of the friction force profile [the time during (Step-1), (Step-2), and (Step3)] of each participant to the average time of all participants.
\item[$\left< 2 \right>$] Average the profiles of all participants per walking speed.
\item[$\left< 3 \right>$] Adjust the impulse by the backward friction force ($B$) and the impulse by the forward friction force ($F$).
\item[$\left< 4 \right>$] Convert to the triangular profile according to the modified impulses ($B'$) and ($F'$), as well as the peak timing of the average profile.
\item[$\left< 5 \right>$] Scale down the triangular profile so that it can be presented by the proposed display.
\end{description}
\vspace{-2.5mm}

{\color{blue}First, to average profiles} of each participant's friction forces applied during walking, we calculated the average time from (Step-1) to (Step-3) and aligned these to the average time $\left< 1 \right>$.
Next, we averaged the profiles {\color{blue}of} each walking speed $\left< 2 \right>$.

\begin{figure}[t]
  \centering
  \includegraphics[width=3.3in]{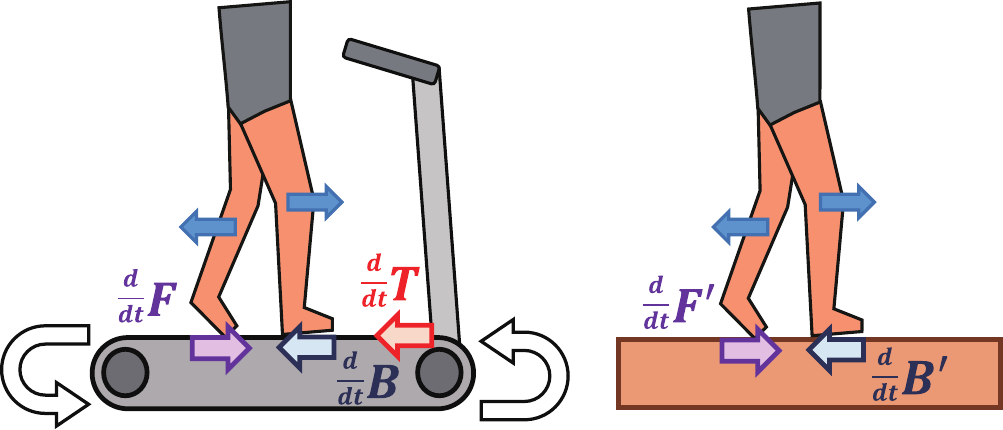} \vspace{-3mm}
  \caption{The impulses of the friction force during walking on the treadmill and over the ground \label{TreadForce}} \vspace{-3mm}
\end{figure}

There is one problem{\color{red}:} the friction force profiles during walking were measured on {\color{blue}a} treadmill, and {\color{blue}these} profiles will be {\color{blue}different} from the profiles during walking over ground.
{\color{blue}Thus,} backward friction force {\color{blue}was measured larger than expected} and the forward friction force {\color{blue}was measured smaller than expected} 
because the treadmill makes the floor belt move backward.
Thus, some adjustments {\color{blue}were} necessary.
As shown in Fig.~\ref{TreadForce}, we named the impulses of 
the backward friction force and the forward friction force measured on the treadmill as $B$ and $F$ respectively. 
Also, we named the impulses by the treadmill as $T$.
{\color{blue}Also}, we named the impulses of the backward friction force and the forward friction force during walking over the ground as $B'$ and $F'$, respectively. These impulses correspond to the impulses we {\color{blue}aim} to represent {\color{blue}using the proposed device}. 
We defined the impulses as the area surrounded by the friction force profiles (Fig.~\ref{Calculation}).
Under this determination, if we suppose that the equations $B=B'+T$ and $F=F'-T$ hold, 
the equation $B'+F'=B+F$ also holds.
{\color{blue}Also}, if we suppose that the body movement during walking at low speed stops {\color{blue}step-by-step}, 
it is considered that the brake impulse $B'$ cancels the driving impulse $F'$.
Assuming that the equation $B'=F'$ holds, the equation $B'=F'=\frac{B+F}{2}$ also holds.
Thus, we scaled the entire friction force profiles to convert 
the surrounded areas to $\frac{B+F}{2}$ $\left< 3 \right>$.

%

{\color{blue}Although} the proposed display can {\color{blue}rapidly} present the friction force, the presented friction force 
becomes a little duller than the output signal.
Thus, it is difficult to present the same force profile measured during walking.
{\color{blue}Also, although} the measured friction force differed for each part of the sole,
the proposed display {\color{blue}applies} the friction force {\color{blue}over} the entire sole.
Thus, we have to convert the force profiles calculated above to {\color{blue}those} that the proposed display can present.
We selected the triangular profile as the output signal because it can be {\color{blue}stably} presented by the proposed device.
{\color{blue}First,} we output the triangular profile corresponding to the backward impulse, 
{\color{blue}followed by }the forward impulse.
The triangular profile reaches the top vertex at the same {\color{blue}time} as the peak timing of the averaged profile.
We regarded the timing of (Step-1) and (Step-3) as the peak timing.
As mentioned above, when the walking speed increased, the state (Step-2) 
{\color{blue}in which the forward friction forces are applied to both the thenar and heel sides simultaneously disappears}. 
Thus, we did not take the timing of (Step-2) into consideration, because we could not interpolate 
the timing between each walking speed based on (Step-2).
The heights of each triangle were calculated so that {\color{blue}their} areas  correspond to the 
impulses{\color{blue},} as mentioned above $\left< 4 \right>$. 

{\color{blue}By this approach}, we could {\color{blue}obtain} the target profiles of the friction force.
However, the current version of the proposed display cannot present as {\color{blue}large a} friction force 
as the target profile. 
Thus, we calculated {\color{blue}a} scaling rate that {\color{blue}scales} down the maximum friction force of 
the target profiles to the maximum friction force that the proposed display can present, and scaled down {\color{red}all the} target profiles by this reduction rate $\left< 5 \right>$.
{\color{blue}Thus, we obtained} the triangular friction force profile per walking speed {\color{blue} and} interpolated the peak timing and the peak friction force between the walking speeds.
Finally, we calculated the actual output signal based on the {\color{blue}relationship} between the 
output signal to the proposed display and the friction force that the proposed display presents.
{\color{blue}To use the value at the time of stable output as the standard value, we calculated this relationship  
based on the measurement result when the rise and fall times of the output signal were set to 0.2~s, where we increased  and decreased the output value linearly to reach the maximum and minimum value in 0.2~s, respectively, with 0.5~s interval.}

\section{Representation of Walking Sensation}
In this section, we investigated whether the proposed display can {\color{blue}realistically represent the walking 
sensation.} 

\subsection{Walking Experience System}
To represent the walking sensation, we developed a system in which 
the output signal is sent to the proposed display according to {\color{blue}an} avatar's walking motion and speed 
in a virtual world. In this system, the graphical images displayed in {\color{blue}the} HMD are rendered according 
to the movement of the avatar's head position.

In the virtual world, the avatar's movement is controlled {\color{blue}using} a game pad (Rumblepad 2, by Logicool Co., Ltd.). This system displays the {\color{blue}view from a} camera placed at the avatar's head position {\color{blue}onto} 
the HMD (Oculus Rift DK2, by Oculus VR, Inc.). We developed this system based on {\color{blue}a} game engine 
(Unity, by Unity Technologies Inc.) and used the character (Ethan) involved in the standard assets 
of Unity as the avatar. Ethan has {\color{blue}a walking animation that was captured using} {\color{red}a} motion capture system. In our system, we utilized {\color{red}this} default animation to display the motion of 
{\color{blue}the} legs during walking. {\color{red}In the measurement test, the friction force {\color{blue}}began} to work at the {\color{blue}moment} the Y-axis' force is confirmed. Thus, we presented the friction force according to the avatar's walking speed 
{\color{red} when Ethan's foot grounded.}
To realize the natural walking sensation, we played the sound of {\color{blue}footsteps binaurally using the Unity plugin}
(3DCeption, by Two Big Ears Ltd.){\color{blue}, corresponding to when the avatar's feet were grounded}.
The sound was played with {\color{red}noise-canceling headphones} (QuietComfort 15, by Bose Corporation) 
to reduce the effects of the environmental noise {\color{blue}, as well as the noise from the motor}.
The {\color{blue}output signal was calculated at a rate of} 1000 times per second.

\subsection{Evaluation of {\color{red}realism}}
\subsubsection{Experimental Setup}
We evaluated the {\color{red}realism} of the walking sensation represented by the proposed display.
{\color{red}We compared our friction force stimuli with both vibration and none stimuli.}
Twelve participants (average age {\color{blue}$23.3\pm1.7$ years})
{\color{red}, including the four participants who partook in
the friction force measurements, joined this experiment.}
Before the experiment, we briefly explained the concept of our device and system.
{\color{blue}Using a questionnaire}, the participants evaluated the {\color{red}realism} of various sensations related to walking 
while controlling the avatar and walking in a virtual space. 
{\color{blue}For} the magnitude estimation, we instructed the participants to  walk before {\color{blue}taking part in} this 
experiment. The {\color{red}realism} of each sensation in the virtual world was scored 
from 0 to 100 points {\color{red}without a visual analog scale} relative to real walking (regarded as 100 points).
{\color{blue}Based on~\cite{Pedal}, we} selected the sensations listed in {\color{red}Table}~\ref{RealitySense} as the evaluation items
while referring to the experiment in {\color{red}the paper, in which the realism of the walking sensation while the user remaining seated was evaluated like our experiment.} 
{\color{red}These items were presented to the participants in both Japanese and English.}

\begin{table}[!b]
\begin{center} \vspace{-4mm}
 \caption{The evaluation items of the {\color{red}realism} of the walking sensation} 
 \vspace{0mm} 
 \begin{tabular}{|p{6.3em}||p{17.8em}|}\hline
Name & Explanation \\ \hline \hline
Driving force & Sensation of the applied driving force to move the body forward \\ \hline
Periodicity & Sensation of the repetitiveness of the footstep sensation \\ \hline
Muscle tension & Sensation of the muscle tension during walking \\ \hline
Kinesthetic sensation & Sensation of the movement of the leg when walking \\ \hline
Brake & Sensation of the applied brake force when the foot grounded \\ \hline
Advancement & Sensation of the body advancing forward \\ \hline
Contact & Sensation of the skin when the foot grounded \\ \hline
Balance control & Sensation of balancing body posture during walking \\ \hline
Hardness & Sensation of the hardness of the ground when the foot is landed\\ \hline 
Texture & Sensation of the texture of the ground when the foot is landed \\ \hline
Timing & Timing in feeling the footstep sensation \\ \hline
Total & Reality 
of the total system \\ \hline
 \end{tabular}
 \label{RealitySense}
\end{center}
\end{table}

\begin{figure}[tb]
  \centering
  \includegraphics[width=3.4in]{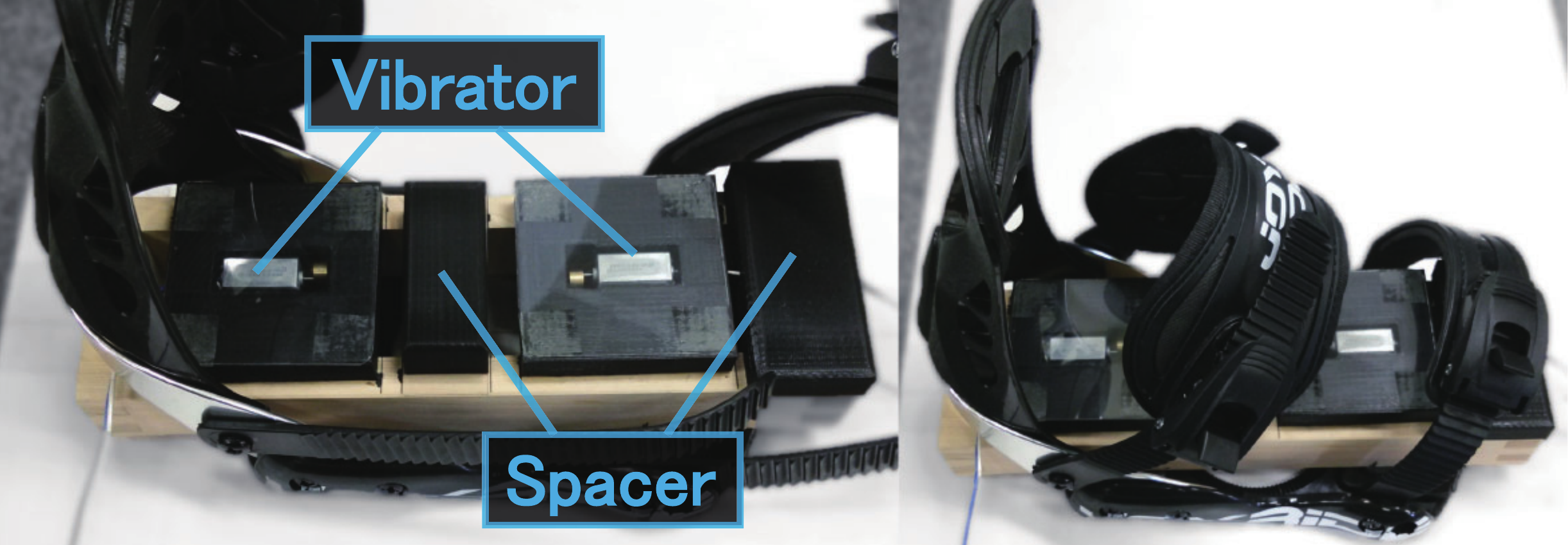} \vspace{-3mm}
  \caption{The outline of VibStep that presents the vibration stimuli\label{VibrationDevice}} \vspace{-3mm}
\end{figure}

{\color{blue}The} participants evaluated the {\color{red}realism} of the walking sensation 
not only {\color{blue}using the proposed display (Friction condition) and} without stimulus (None condition). 
{\color{blue}Also, the participants} evaluated the {\color{red}realism} of a vibration stimulus (Vibration condition) that can be presented {\color{blue}using} a smaller device than the proposed display.
We used 12~V {\color{red}vibrators} (a13081500ux0075, by UXCell) to present the vibration stimuli and, to 
transmit the vibration to the sole, we attached the vibrator embedded plate to the snowboard binding. 
Vibrator-embedded {\color{red}plates} were attached on both the thenar and heel sides. 
We used the snowboard binding (JOBG-340 S/M, by ACTIVE Co., Ltd.) because the binding 
used for the proposed display was already unavailable.
Fig.~\ref{VibrationDevice} shows the device to present the vibration stimuli (tentatively called VibStep).

\begin{figure}[tb]
  \centering
  \includegraphics[width=3.1in]{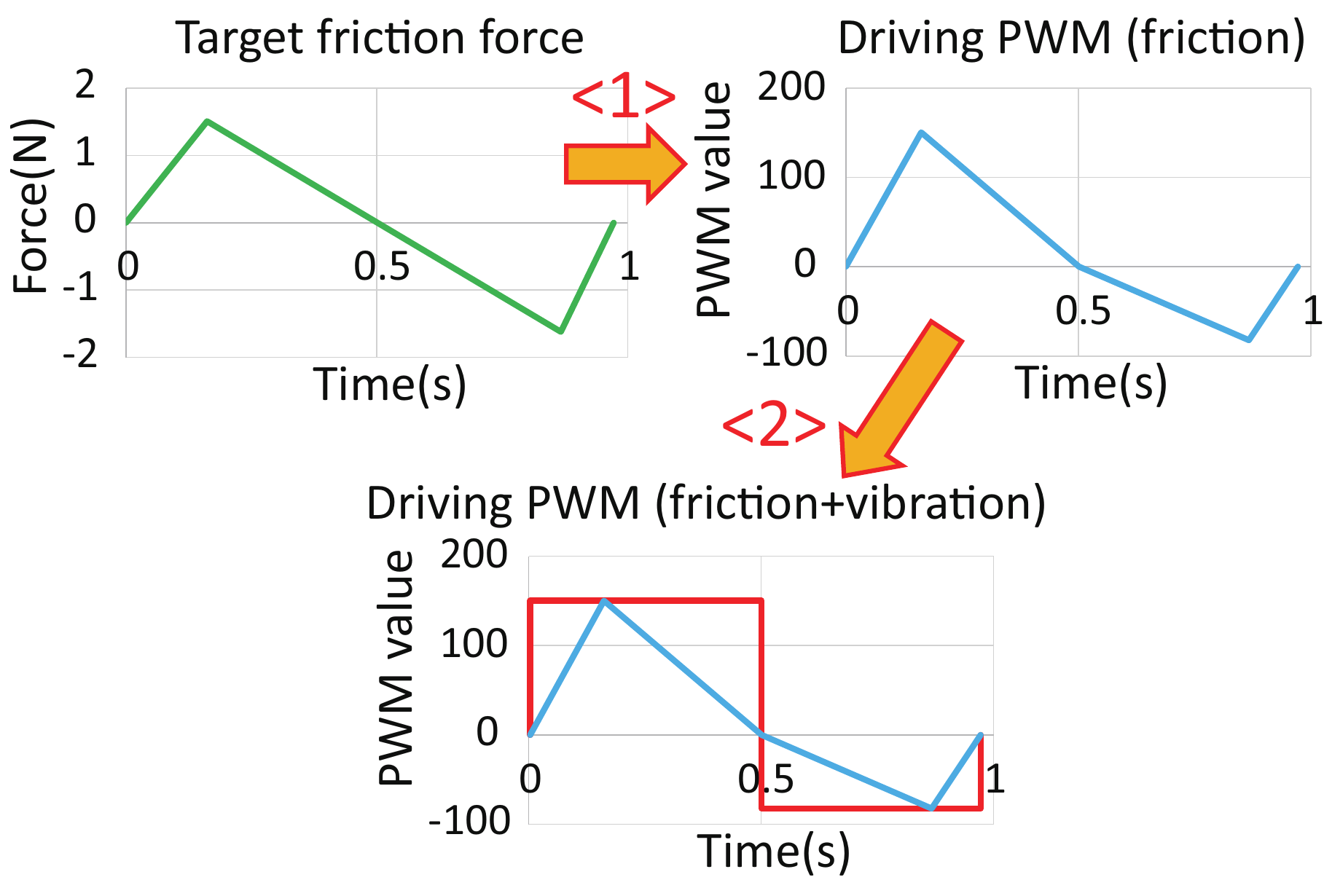} \vspace{-5mm}
  \caption{The procedure to calculate the output signal to VibStep \label{VibrationOutput}} \vspace{-1mm}
\end{figure}

As the output signal to VibStep, we used {\color{blue}a} rectangular signal that covers the 
triangular PWM signal {\color{blue}used} for the proposed device. At the preparation stage of this experiment, we found that 
if we do not {\color{blue}maintain an} output signal to the vibrator, {\color{blue}the user is unable to perceive the stimulus.} 
If we give the triangular output signal to VibStep, the vibration stimulus 
{\color{blue}is insufficient} because the peak load of the triangular signal {\color{blue}is fleeting}. 
Thus, we decided to give a rectangular signal to VibStep.
Fig.~\ref{VibrationOutput} shows {\color{blue}the example of} the specific procedure 
{\color{blue}used} to calculate the output signal to VibStep. After we converted the target friction force profile 
to the triangular PWM signal to the proposed device $\left< 1 \right>$, we calculated the rectangular PWM 
signal that covers the triangular signal $\left< 2 \right>$.
Unlike the proposed display, the stimulus display part of VibStep is divided into the thenar  and heel sides. 
When the output signal is {\color{blue}positive}, we output it to the vibrator on the heel side; 
when the output signal is {\color{blue}negative}, we output it to the vibrator on the thenar side.
This type of vibration output presenting on the thenar side following the 
presentation on the heel side {\color{blue}was also adopted in a previous study}~\cite{Vibration}.


\begin{figure}[tb]
  \centering
  \includegraphics[height=1.3in]{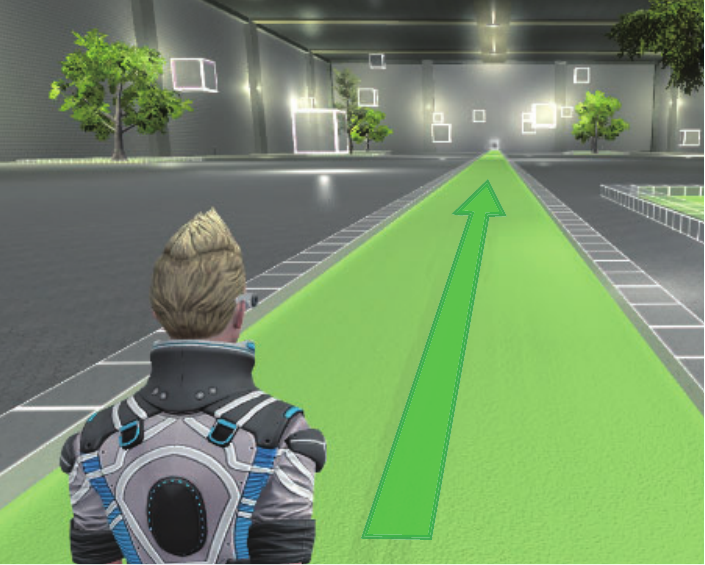}
  \includegraphics[height=1.3in]{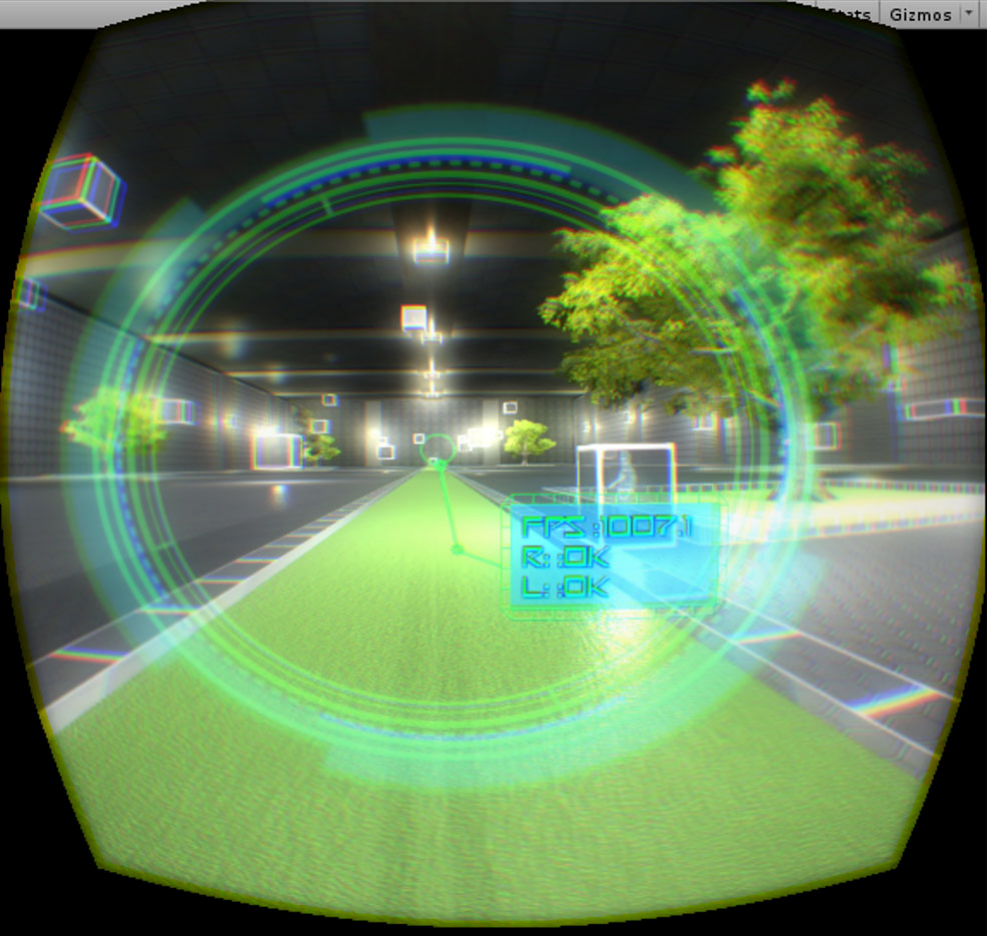}\\
  \hspace{7mm}(a)\hspace{35mm}(b) \vspace{-3mm}
  \caption{The experimental setup for the evaluation of the walking sensation (a) virtual space for the experiment (b) an example of the image displayed in HMD \label{RealityTest}} \vspace{-5mm}
\end{figure}

At the beginning of this experiment, the participants evaluated the {\color{red}realism} on the {\color{blue}'None'} condition.
Next, they evaluated the {\color{red}realism} on the 'Friction' condition or the 'Vibration' condition.
Finally, they evaluated the {\color{red}realism} on another condition which they did not evaluated in the second evaluation.
In the second evaluation, {\color{red}half of the} participants evaluated on the 'Friction' condition, 
and the other half of the participants evaluated on the 'Vibration' condition.
In our system, the presented friction force and the graphical images are changed 
according to the avatar's walking speed. However, in this experiment, 
we limited the walking speed to 
{\color{blue}1.0, 2.5, and 4.0 km/h 
(at which we measured the friction force during actually walking).}
{\color{blue}For} each stimulus condition (None, Vibration, {\color{blue}and} Friction), 
{\color{blue} half of the participants evaluated the {\color{red}realism} in the order of 1.0, 2.5}, and 4.0 km/h, 
and the other half evaluated {\color{blue}the realism in the order of 4.0, 2.5}, and 1.0 km/h.

The participants evaluated the {\color{red}realism} {\color{blue}of} each condition by controlling the avatar {\color{blue}using a} gamepad. In this experiment, we limited the avatar's movement {\color{blue}to a simple} forward movement.
When the participants tilted the joystick to a certain extent, the avatar started walking 
at the specified speed. Fig.~\ref{RealityTest} shows the experimental set up of 
the virtual space. To make it easy to sense the distance, we used {\color{blue}a} scene in which 
some objects are placed.

\begin{figure}[tb]
  \centering 
  \includegraphics[width=2.75in]{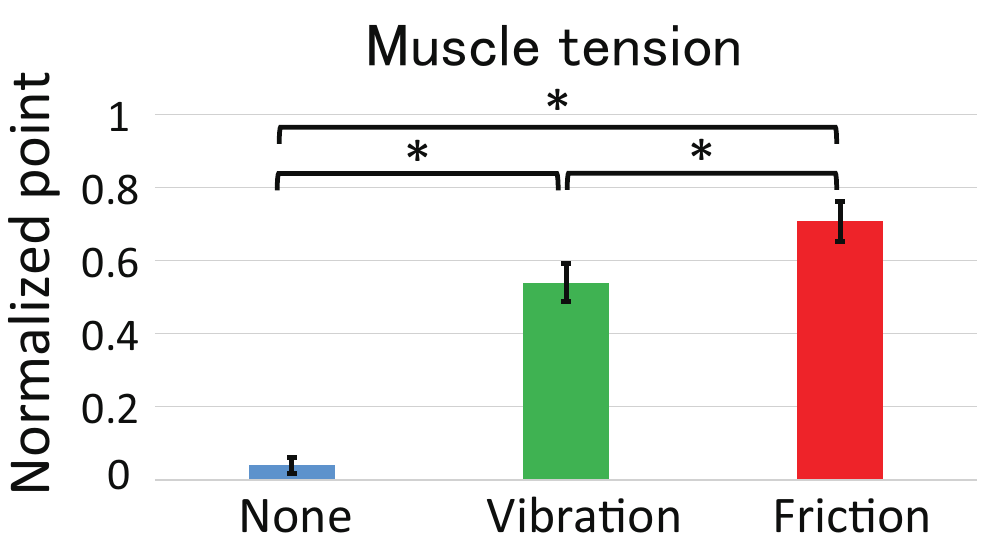}\\ \vspace{0.5mm} 
  \includegraphics[width=2.75in]{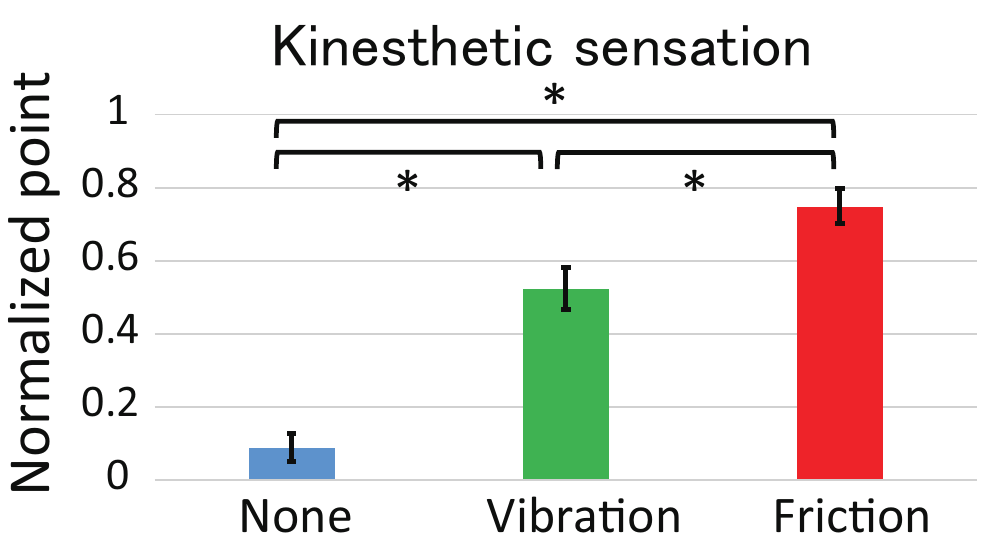}\\ \vspace{0.5mm}
  \includegraphics[width=2.75in]{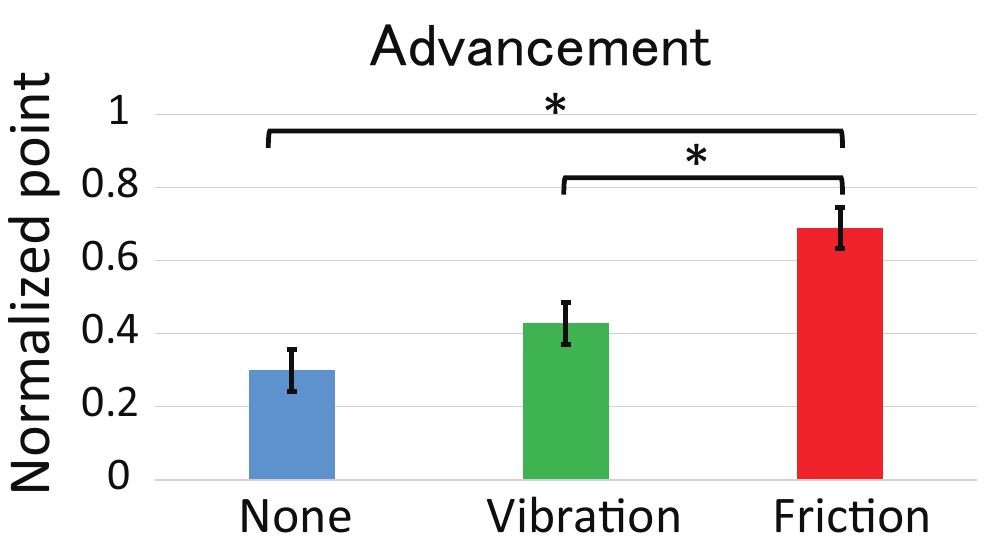} \vspace{-4mm}
  \caption{The results of the {\color{red}realism} evaluation of the walking sensation (excerpt) \label{RealityResult}} \vspace{-5mm}
\end{figure}

\subsubsection{Results}
{\color{red}
We conducted Friedman test for speed and stimuli conditions with repeated measures, which does not assume normality of the samples' distribution. 
We also conducted multiple {\color{blue}comparisons} by the post hoc test using Wilcoxon rank sum test with Bonferroni's correction to the items that we could not confirm a main effect of speed in Friedman analysis and ignore the speed difference of the samples. 
There were a large variance between participants.
Also, for each item, the} maximum and minimum {\color{blue}points} of each participant 
were {\color{blue}the} same only when both {\color{blue}points} were 0. 
Thus, as shown below, we normalized the {\color{blue}points} of each participant 
before averaging the points of all participants.

\begin{enumerate}\vspace{-1.5mm}
\item For each participant, {\color{blue}we} calculated the maximum point $x_{\rm max}$ and the minimum point $x_{\rm min}$ of all data ({\color{blue}three} stimuli $\times$ three speeds) belonging to the evaluation item $x$ \vspace{-1.25mm}
\item For each participant, {\color{blue}we} normalized each data $n$ belonging to the evaluation item $x$ using 
the equation $n_{\rm new} = (n-x_{\rm min}) / (x_{\rm max}-x_{\rm min})$ from 0 to 1 \vspace{-1.25mm}
\item {\color{blue}When} $x_{\rm max}$ and $x_{\rm min}$ are {\color{blue}at} the same point, $n_{\rm new} = 0$ \vspace{-1.5mm}
\end{enumerate}

Fig.~\ref{RealityResult} shows {\color{red}the averaged normalized points of} {\color{blue}those items for} which we {\color{blue}found} that the 'Friction' condition was 
significantly more real than the 'Vibration' condition (p$<$0.05). 
{\color{blue}The 'Vibration' and the 'None' conditions outscored the 'Friction' condition for no items.} 

\subsubsection{{\color{red}Discussion}}
\begin{figure}[tb]
  \centering 
  \includegraphics[width=2.75in]{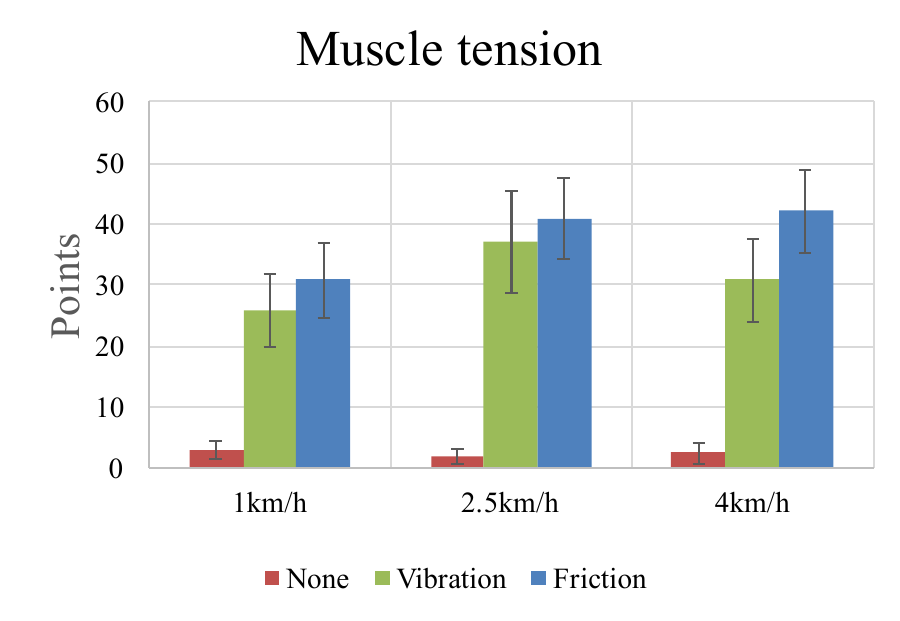}\\ \vspace{0.5mm} 
  \includegraphics[width=2.75in]{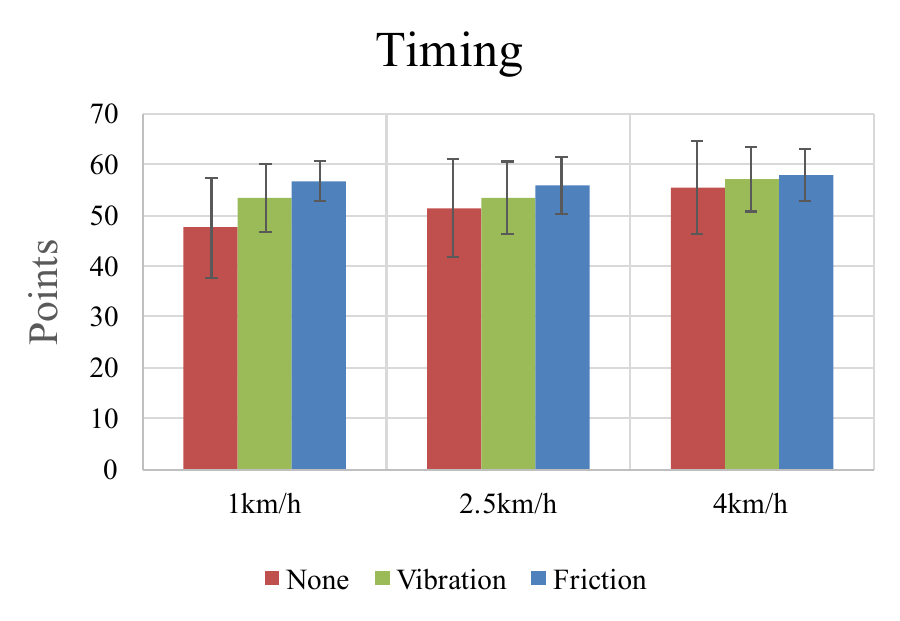} \vspace{-4mm}
  \caption{{\color{red}The results of raw points of the realism evaluation of the walking sensation before normalization (excerpt) \label{RealityResultRaw}}} \vspace{-5mm}
\end{figure}

{\color{blue}From these findings, we were able to confirm that 
``muscle tension", ``kinesthetic sensation", and ``advancement" felt more real using the proposed display than when using a vibration stimulus}.
We {\color{blue}believe} that the stimulus presented by the proposed display generated not only the sensation on the sole
of the foot but also the illusion of the entire leg's motion. {\color{blue}This is based on the detection of significant 
differences for the two items (``muscle tension" and ``kinesthetic sensation"), which are} related to the entire leg.
{\color{blue}Also}, the improvement {\color{blue}in the ``advancement" sensation} was {\color{blue}greater than for} other items.
This indicates that the representation of the friction force when kicking the ground generates {\color{blue}a} 
strong advancing feeling.
Fig.~\ref{RealityResultRaw} shows the representative raw points before normalization.

{\color{blue}However, we detected no significant difference in
the {\color{red}realism} of the friction force when kicking the ground (the item of ``driving force"). }
In the current version of the proposed display, we presented the friction force on the entire sole by 
{\color{blue}the} 1-DOF mechanism. Thus, unlike the vibration stimulus {\color{blue}of} VibStep, the proposed display cannot 
represent the shift {\color{blue}in} the load position from the heel side to the thenar side during walking{\color{blue}, which might limit its {\color{red}realism}}. 
If we increase the degrees of freedom of the proposed display and represent the weight shift, 
{\color{blue}the {\color{red}realism} would likely be 
improved.}

{\color{red}
Example comments from the participants are ``Fixing the ankle reduces the advancement a bit`` and ``I really felt the walking sensation when the friction force was displayed at the fastest speed`` while another participant reported ``The presented friction force at the fastest speed was too strong``. Also, one participant reported a little fatigue due to the VR sickness.
We should mention the lack of proper counterbalancing in the order of stimuli conditions. We always displayed {\color{blue}the 'None'} condition first. 
Thus, {\color{blue}our results might include an} effect caused by adding haptic feedback, {\color{blue}rather than} removing haptic feedback.  However, the influence {\color{blue}of the order effect} would be limited, and would not reverse the differences between {\color{blue}the 'None' and 'Friction' conditions (Fig.~\ref{RealityResult}). }
}

\section{Conclusion}
{\color{green}In this paper we describe} {\color{blue}the development of a novel haptic device} to represent the walking sensation with a 
minimal mechanism and without fatigue. 
The proposed display represents the reaction force generated when the 
{\color{blue}user's} foot is landed and when {\color{blue}kicking} the ground by representing the longitudinal friction force applied on the sole during walking.
From the measurement test, {\color{blue}although} there were individual differences in the presented force, we {\color{blue}were able to} control the friction force. 
{\color{blue}Also, we found that 
the proposed display could present rapid stimuli within about 0.1 s}.

Based on the measured friction force applied on the sole during {\color{blue}real} 
walking, we presented the friction force according to the avatar's 
walking motion in a virtual world. As a result, we succeeded {\color{blue}in achieving 
a} significantly more real advancing feeling and kinesthetic sensation 
than the vibration stimuli. 
{\color{blue}However, we detected no} significant improvement {\color{blue}in} the {\color{red}realism} of some {\color{blue}sensations 
related to walking}. 
{\color{red}When we calculated the 
output signal to HapStep, we simplified and scaled down the targeted friction force due to the limited performance of our device. 
{\color{blue}If a device is developed that can 
present larger friction forces more accurately, more realistic walking sensation might} be realized. 
{\color{blue}On the other hand, as mentioned above, some users expressed that ``the presented friction force was too strong".} 
The integration of the proposed device with the vibrotactile one, which can modulate stimuli widely, would be good choice to enhance the 'texture' and 'timing' representation. If we attempt to represent a compliant ground rather than solid ground, we should represent the effects caused by the deformation of a shoe sole, e.g. stretching force applied on the foot.}  In our measurements during walking, we cannot truly isolate the gravitational force from the longitudinal friction force. Thus, it is possible that the load {\color{blue}measured} friction force was larger than the actual load. {\color{blue}Therefore, the measurement method will need to be improved as a part of future studies.}

The current version of the proposed display tries to represent only the 
friction force applied when the foot is landed, and 
the walker kicks the ground. However, there is a possibility 
that the proposed display can represent other moving sensations, such as 
jumping or sliding on ice. 
If we can represent these additional moving sensations, it will become possible to achieve a 
richer virtual experience.



\end{document}